\documentclass[reprint,superscriptaddress,preprintnumbers,nofootinbib,amsmath,amssymb,aps,prd,showkeys,showpacs]{revtex4-2}

\usepackage{graphicx}% Include figure files
\usepackage{dcolumn}% Align table columns on decimal point
\usepackage{bm}% bold math
\usepackage{array} % 导言区引入
\newcolumntype{M}[1]{>{\centering\arraybackslash}m{#1}}
\usepackage{endnotes} 
\usepackage{adjustbox}

\usepackage{amssymb}    %Allows use math symbols
\usepackage{color}         %Allows {\color{red} Hello World} or \colorbox{blue}{Hello World}
\usepackage{graphicx}     %Allows\includegraphics{figure.pdf}
\usepackage{mathrsfs} %Allows \mathscr{HELLO}
\usepackage{hyperref} %Automatically links \label and \ref commands; Always load last
\hypersetup{colorlinks=true,linkcolor=blue,citecolor=blue}
\usepackage{ upgreek } % for uptau
\usepackage{color,xcolor,fancybox,epsf,rotating,colordvi}
\usepackage{booktabs}
\usepackage{multirow}

\usepackage{soul}
\setstcolor{red}
\usepackage{cancel}

\newcommand{\tanb}{\tan\!\beta}

\newcommand{\abm}{{~\rm ab}^{-1}}
\begin{document}

% \preprint{APS/123-QED}
\title{Particle-level transformers for 95 GeV Higgs boson searches at future $e^+e^-$ Higgs factories}

\author{Yabo Dong}
% \email[]{dongyb@henu.edu.cn}
\affiliation{School of Physics and Electronics, Henan University, Kaifeng 475004, China}

\author{Manqi Ruan}
% \email[]{manqi.ruan@ihep.ac.cn}
\affiliation{Institute of High Energy Physics, Chinese Academy of Sciences, Beijing 100049, China}

\author{Kun Wang}
% \email[]{kwang@usst.edu.cn}
\email[Corresponding author:]{kwang@usst.edu.cn} 
\affiliation{College of Science, University of Shanghai for Science and Technology, Shanghai 200093, China}

\author{Haijun Yang}
% \email[]{haijun.yang@sjtu.edu.cn}
\affiliation{State Key Laboratory of Dark Matter Physics, Key Laboratory for Particle Astrophysics and Cosmology (MOE), Shanghai Key Laboratory for Particle Physics and Cosmology (SKLPPC),  School of Physics and Astronomy \mbox{\normalfont\&} Tsung-Dao Lee Institute, Shanghai Jiao Tong University, Shanghai 200240}

\author{Jingya Zhu}
% \email[]{zhujy@henu.edu.cn} 
\email[Corresponding author:]{zhujy@henu.edu.cn} 
\affiliation{School of Physics and Electronics, Henan University, Kaifeng 475004, China}

% \collaboration{CLEO Collaboration}%\noaffiliation

\date{\today}% It is always \today, today,

\begin{abstract}
Motivated by several mild excesses around 95~GeV, we investigate the prospects for a light scalar $S$ produced via Higgsstrahlung, $e^+e^- \to Z(\mu^+\mu^-)S$, at future $e^+e^-$ Higgs factories. We take the CEPC as a benchmark, with a center-of-mass energy of $\sqrt{s}=240$ GeV and an integrated luminosity of $L=20~\mathrm{ab}^{-1}$. We focus on the decay modes $S\to\tau^+\tau^-$ and $S\to b\bar b$. To maximize sensitivity, we employ the particle-level transformer networks Particle Transformer (ParT) and its more-interactive variant MIParT, which exploit the features of all reconstructed objects and their correlations. 
For a representative signal benchmark, this approach improves the expected statistical precision on the signal strength by factors of 2.3 in the $\tau^+\tau^-$ channel and 1.4 in the $b\bar b$ channel compared to a cut-based analysis. Within the flipped Next-to-Two-Higgs-Doublet Model (N2HDM-F), the CEPC can measure the signal strength with a statistical precision down to 1.0\% in the $\tau^+\tau^-$ channel and 0.69\% in the $b\bar b$ channel using MIParT. It can achieve a $5\sigma$ discovery for $\mu_{\tau\tau}^{ZS}>1.6\times10^{-2}$ or $\mu_{bb}^{ZS}>5.0\times10^{-3}$, and reach 1\% precision for $\mu_{\tau\tau}^{ZS}>0.93$ or $\mu_{bb}^{ZS}>0.14$. These gains are expected to qualitatively carry over to other future lepton colliders such as FCC-ee and the ILC. Our results demonstrate the potential of particle-level machine-learning techniques to strengthen light Higgs searches at future $e^+e^-$ Higgs factories.
\end{abstract}

\maketitle
\newpage

% \tableofcontents
% \newpage

%%%

\section{Introduction}
\label{sec:intro}

The discovery of the Higgs boson at the LHC in 2012 \cite{CMS:2012qbp, ATLAS:2012yve} completed the last missing piece of the Standard Model (SM) puzzle. 
However, several unresolved questions continue to cast shadows over the SM, among which the origin of the baryon asymmetry of the universe remains one of the most compelling \cite{Kajantie:1995kf, Farrar:1993sp, Gavela:1993ts}. 
According to the Sakharov conditions \cite{Sakharov:1967dj}, three criteria must be fulfilled to generate the observed matter–antimatter asymmetry: 
(i) baryon number violation, 
(ii) sufficient C and CP violation, and 
(iii) departure from thermal equilibrium. 
In the SM, the C and CP violations are too weak, and the electroweak phase transition (EWPT) associated with the 125~GeV Higgs boson is merely a smooth crossover rather than a strong first-order transition \cite{Huet:1994jb, Rummukainen:1998nu}. 
These shortcomings strongly suggest that new physics beyond the SM is required to account for the baryon asymmetry. 
A minimal yet well-motivated approach is to extend the Higgs sector by introducing additional scalar fields. 
Such extensions can naturally provide new sources of C and CP violation and can also induce a strong first-order EWPT, thereby realizing the necessary out-of-equilibrium conditions \cite{Anisha:2022hgv}.

Several experimental excesses near 95 GeV have been reported across different search channels, suggesting intriguing hints of a light scalar resonance. 
The earliest indication dates back to 2003, when the LEP experiment observed a local significance of $2.3\sigma$ for a hypothetical lighter Higgs boson $S$ with a mass around 98 GeV in the process $e^+e^- \rightarrow ZS\,(\rightarrow b\bar{b})$~\cite{LEPWorkingGroupforHiggsbosonsearches:2003ing}. 
In 2018, the CMS Collaboration reported a local excess of $2.8\sigma$ in the $gg \to S \to \gamma\gamma$ channel at a mass of 95.3~GeV, based on the combined 8~TeV (19.7~fb$^{-1}$) and 13~TeV (35.9~fb$^{-1}$) datasets~\cite{CMS:2018cyk}. 
Additional evidence emerged in 2022, when CMS reported a $3.1\sigma$ excess in the $pp \to S \to \tau^+\tau^-$ channel near 100~GeV~\cite{CMS:2022goy}. 
More recently, in 2024, both major LHC collaborations released updated diphoton results: CMS observed a $2.9\sigma$ excess at 95.4~GeV~\cite{CMS:2024yhz}, while ATLAS reported a milder $1.7\sigma$ excess at 95.3~GeV in $pp \to S \to \gamma\gamma$~\cite{ATLAS:2024bjr}. 
Taken together, these observations may point toward the existence of a light Higgs with a mass around 95.5 GeV. 

The collider phenomenology of the hypothetical light Higgs boson has been extensively explored in recent studies through Monte Carlo (MC) simulations~\cite{Sharma:2024vhv, Wang:2024bkg, Dutta:2023cig, Dong:2024ipo, Dong:2025orv}. 
Hadron colliders, such as the Large Hadron Collider (LHC), operate at high center-of-mass energies, enabling the exploration of a wide range of physical processes~\cite{CMS:2007sch, ATLAS:2010ojh}. 
However, the complex QCD background at hadron colliders poses a major challenge for precision measurements and for probing subtle new-physics signals. 
Our previous analysis indicates that achieving a $5\sigma$ discovery significance in the top-pair-associated diphoton channel, $pp \to t\bar{t}S(\to \gamma\gamma)$, at the High-Luminosity LHC (HL-LHC) requires a production cross section exceeding approximately 0.3~fb~\cite{Dong:2024ipo}. 
Signals with smaller cross sections may instead be probed through alternative decay channels or at future collider facilities. 

Future Higgs factories, such as the Future Circular Collider (FCC-ee)~\cite{FCC:2018evy, FCC:2018byv}, the International Linear Collider (ILC)~\cite{ILC:2013jhg, Behnke:2013xla, Asner:2013psa}, and the Circular Electron-Positron Collider (CEPC)~\cite{CEPCStudyGroup:2018ghi, CEPCStudyGroup:2018rmc, An:2018dwb, CEPCStudyGroup:2023quu}, provide a promising avenue for probing a possible 95~GeV light Higgs boson and measuring its couplings with high precision. 
This enhanced sensitivity stems from the much cleaner environment of lepton colliders compared to hadron colliders.
Our study focuses on the CEPC, motivated by its suitable center-of-mass energy and high integrated luminosity. 
The scenarios of the other two colliders will be discussed at the end of the paper.
The CEPC aims to study the properties of the 125~GeV Higgs boson with unprecedented accuracy and is expected to deliver over 20~ab$^{-1}$ at 240~GeV~\cite{CEPCStudyGroup:2018ghi, CEPCStudyGroup:2018rmc, An:2018dwb, CEPCStudyGroup:2023quu}. 
Given that the dominant production mechanism for a 95~GeV scalar at the CEPC is the Higgsstrahlung process ($e^+e^- \to ZS$), this work is dedicated to its detailed investigation.

Our previous work~\cite{Dong:2025orv} demonstrated that machine learning (ML) algorithms, including eXtreme Gradient Boosting (XGBoost), Gradient Boosting Decision Trees (GBDT), and deep neural networks (DNN), can enhance the sensitivity of searches for a light Higgs boson around 95 GeV.
Furthermore, in recent years, several advanced ML architectures have shown considerable promise in jet tagging tasks~\cite{Komiske:2018cqr, Qu:2019gqs, Mikuni:2020wpr, Mikuni:2021pou, Gong:2022lye, Qu:2022mxj, Wu:2024thh}. 
Among these, transformer-based models such as the Particle Transformer (ParT)~\cite{Qu:2022mxj} and the More-Interaction Particle Transformer (MIParT)~\cite{Wu:2024thh} have demonstrated outstanding performance. 
Their strong performance stems primarily from the attention mechanism, which effectively captures complex correlations among final-state particles and yields a more expressive representation of jet substructure. This capability precisely addresses a key shortcoming of simpler network architectures, such as DNN.
Overall, particle-level ML techniques offer substantial potential for new-physics searches and are expected to significantly improve both sensitivity and measurement precision at present and future colliders.

The persistent experimental anomalies near 95 GeV have motivated theoretical interpretations.
These interpretations can be broadly classified according to the decay channels considered: focus solely on the $\gamma\gamma$ channel~\cite{Kundu:2019nqo, Liu:2018ryo, Borah:2023hqw,  Abdelalim:2020xfk, Ashanujjaman:2023etj, Banik:2023ecr, Dong:2024ipo, Coloretti:2023yyq, Maniatis:2023aww, Wang:2018vxp, Cao:2019ofo, Li:2023kbf, Wang:2022okq}, analyse both $\gamma\gamma$ and $b\bar{b}$ channels~\cite{Aguilar-Saavedra:2020wrj, Escribano:2023hxj, Dev:2023kzu, Ahriche:2023hho, Biekotter:2019kde, Benbrik:2025hol, Heinemeyer:2021msz, Biekotter:2021ovi, Banik:2024ugs, Dutta:2025nmy, Xu:2025vmy, Benbrik:2022azi, Biekotter:2020cjs, Cao:2016uwt, Li:2022etb, Cao:2023gkc, Lian:2024smg, Cao:2024axg, Ellwanger:2023zjc, Ellwanger:2024vvs, Choi:2019yrv, Ellwanger:2024txc} or $\gamma\gamma$ and $\tau^+\tau^-$ channels~\cite{Ge:2024rdr, YaserAyazi:2024hpj, Chang:2025bjt}, while consider all three channels, $\gamma\gamma$, $b\bar{b}$, and $\tau^+\tau^-$~ \cite{Khanna:2024bah, Biekotter:2021qbc, Aguilar-Saavedra:2023tql, Biekotter:2022jyr, Biekotter:2022abc, Biekotter:2023jld, Biekotter:2023oen, Arcadi:2023smv, Benbrik:2024ptw, Iguro:2022fel, Azevedo:2023zkg, Belyaev:2023xnv, Chen:2023bqr, Mondal:2025tzi, Kundu:2024sip, Hmissou:2025uep, Hmissou:2025riw, Benbrik:2025wkz, Ahriche:2023wkj}. These anomalies have been interpreted through a twin-peak scenario in the Georgi-Machacek Model~\cite{Chen:2023bqr, Ahriche:2023wkj, Du:2025eop}.
The type-II and flipped Next-to-Two-Higgs-Doublet Model scenarios (N2HDM-II and N2HDM-F) provide particularly elegant solutions.
They offer additional scalar degrees of freedom that can simultaneously account for the observed $b\bar{b}$ and $\tau^+ \tau^-$ anomalies~\cite{Biekotter:2019kde, Biekotter:2021qbc, Aguilar-Saavedra:2023tql, Heinemeyer:2021msz, Biekotter:2022jyr, Sassi:2025dyj}.
In this study, we perform a comprehensive scan of the N2HDM parameter space to examine the model's compatibility with the observed 95~GeV excess. 
Subsequently, an MC simulation and a cut-based jet-level analysis are carried out for the Higgsstrahlung process, $e^+e^- \to ZS$, in various final states at the 240~GeV CEPC. 
To further enhance the search sensitivity and measurement precision for the 95~GeV light Higgs boson, we extract features from all final-state particles and employ them to train the ParT and MIParT networks. 
Finally, we present a detailed discussion of the achievable parameter coverage and measurement precision at the CEPC, both with and without the inclusion of machine learning, and demonstrate the model-independent discovery potential of the CEPC.

The remainder of this paper is organized as follows. 
In Sec.~\ref{sec:scan}, we briefly introduce the N2HDM-F and present the results of the parameter-space scan. 
Section~\ref{sec:mc} describes the MC simulation and the cut-based analysis strategy. 
The implementation of the ParT and MIParT networks, together with the corresponding performance studies and discussions, is detailed in Sec.~\ref{sec:MIParT}. 
Finally, the conclusions are summarized in Sec.~\ref{sec:conclusion}.

\section{\label{sec:scan}The broken-phase flipped N2HDM} 

The Next-to-Two-Higgs-Doublet Model (N2HDM) extends the Standard Model scalar sector through the introduction of a second $SU(2)_L$ Higgs doublet $\Phi_2$ 
and an additional real scalar singlet $\Phi_S$~\cite{Chen:2013jvg, Drozd:2014yla,vonBuddenbrock:2016rmr, Muhlleitner:2016mzt}. 
After electroweak symmetry breaking, the model yields three neutral CP-even Higgs states, one of which corresponds to the observed SM-like Higgs boson at 125\,GeV. 
A lighter state can naturally arise around 95\,GeV, providing a potential explanation for the mild excesses reported in several experiments. 
In the following, we identify the light CP-even state $H_1$ with the putative 95 GeV resonance and denote it by $S\equiv H_1$ whenever convenient.

\subsection{Model Setup}
The scalar potential of the N2HDM can be written as~\cite{Muhlleitner:2016mzt}

\begin{widetext}
\begin{equation}
\begin{aligned}
V  = & m_{11}^2|\Phi_1|^2+m_{22}^2|\Phi_2|^2-m_{12}^2(\Phi_1^\dagger\Phi_2+\mathrm{h}.\mathrm{c}.)+\frac{\lambda_1}{2}(\Phi_1^\dagger\Phi_1)^2+\frac{\lambda_2}{2}(\Phi_2^\dagger\Phi_2)^2 \\
&+\lambda_3(\Phi_1^\dagger\Phi_1)(\Phi_2^\dagger\Phi_2)+\lambda_4(\Phi_1^\dagger\Phi_2)(\Phi_2^\dagger\Phi_1)+\frac{\lambda_5}{2}[(\Phi_1^\dagger\Phi_2)^2+\mathrm{h}.\mathrm{c}.] \\
 & +\frac{1}{2}m_{\Phi_S}^2\Phi_S^2+\frac{\lambda_6}{8}\Phi_S^4+\frac{\lambda_7}{2}(\Phi_1^\dagger\Phi_1)\Phi_S^2+\frac{\lambda_8}{2}(\Phi_2^\dagger\Phi_2)\Phi_S^2.
\end{aligned}
\end{equation}
\end{widetext}
The potential respects a softly broken $Z_2$ symmetry ($\Phi_2 \to -\Phi_2$) and an exact $Z_2'$ symmetry ($\Phi_S \to -\Phi_S$), which forbids linear terms in $\Phi_S$ and explains the absence of certain couplings.

After electroweak symmetry breaking, the N2HDM particle spectrum contains three CP-even neutral scalars $H_{1,2,3}$ (The masses of the three $H_{1,2,3}$ increase in sequence), one pseudoscalar $A$, and a pair of charged scalars $H^\pm$.
The electroweak sector is in the broken phase when both doublets and the singlet acquire nonzero vacuum expectation values (VEVs). 
In this scenario, the three CP-even scalar fields mix through a $3\times 3$ rotation matrix, which have the following form~\cite{Muhlleitner:2016mzt}
\begin{widetext}
    \begin{equation}
   R = 
   \begin{pmatrix}
   c_{\alpha_1} c_{\alpha_2} & s_{\alpha_1} c_{\alpha_2} & s_{\alpha_2} \\
   -(c_{\alpha_1} s_{\alpha_2} s_{\alpha_3} + s_{\alpha_1} c_{\alpha_3}) & c_{\alpha_1} c_{\alpha_3} - s_{\alpha_1} s_{\alpha_2} s_{\alpha_3} & c_{\alpha_2} s_{\alpha_3} \\
   -c_{\alpha_1} s_{\alpha_2} c_{\alpha_3} + s_{\alpha_1} s_{\alpha_3} & -(c_{\alpha_1} s_{\alpha_3} + s_{\alpha_1} s_{\alpha_2} c_{\alpha_3}) & c_{\alpha_2} c_{\alpha_3}
   \end{pmatrix},
   \end{equation}
\end{widetext}
where $\mathrm{c}_\alpha \equiv \mathrm{cos}_\alpha$ and $\mathrm{s}_\alpha \equiv \mathrm{sin}_\alpha$. The parameters of N2HDM can then be parameterized as:
\begin{equation}
\begin{aligned}
    m_{H_1},\ m_{H_2},\ m_{H_3},\ m_{H_A},\ m_{H^{\pm}}, \\
    \alpha_1,\ \alpha_2,\ \alpha_3,\ m_{12}^2,\ v,\ v_S,\ \tanb,
\end{aligned}
\end{equation}
where $m_{H_1}$, $m_{H_2}$, $m_{H_3}$, $m_{H_A}$, and $m_{H^{\pm}}$ denote the masses of $H_1$, $H_2$, $H_3$, $A$, and $H^{\pm}$, respectively; $\alpha_1$, $\alpha_2$, and $\alpha_3$ are the mixing angles of the three CP-even neutral scalars; $m_{12}^2$ is the soft $Z_2$-breaking parameter; $v=246$ GeV and $v_S$ are the VEVs of the electroweak doublet and the singlet field, respectively; and $\tanb$ is the ratio of the two doublet VEVs.

The most interesting part of the couplings are the CP-even Higgs $H_i$ to the massive gauge bosons, the quarks, and the leptons. 
The reduced coupling of Higgs $H_i$ to the massive gauge bosons $V=W,\ Z$ ($C_{H_iVV}$) can be given by 
\begin{equation}
    C_{H_iVV} = 
\begin{cases}
c_{\alpha_2}c_{\beta-\alpha_1},& i=1, \\
-c_{\beta-\alpha_1}s_{\alpha_2}s_{\alpha_3}+c_{\alpha_3}s_{\beta-\alpha_1} ,& i=2,  \\
-c_{\beta-\alpha_1}s_{\alpha_2}c_{\alpha_3}-s_{\alpha_3}s_{\beta-\alpha_1} ,& i=3, 
\end{cases}
\end{equation}
where $c_x$ denotes $\mathrm{cos}(x)$ and $s_x$ denotes $\mathrm{sin}(x)$ for the angles $x=\alpha_1,\ \alpha_2,\ \alpha_3,\ \mathrm{and}\ \beta-\alpha_1$.

In the N2HDM-F scenario, the reduced coupling of Higgs $H_i$ to up-type quarks and leptons $f$ ($C_{H_iff}$) can be given by
\begin{equation}
    C_{H_iff} = 
\begin{cases}
s_{\alpha_1}c_{\alpha_2}/s_{\beta},& i=1, \\
(c_{\alpha_1}c_{\alpha_3}-s_{\alpha_1}s_{\alpha_2}s_{\alpha_3})/s_{\beta} ,& i=2,  \\
-(c_{\alpha_1}s_{\alpha_3}+s_{\alpha_1}s_{\alpha_2}c_{\alpha_3})/s_{\beta} ,& i=3. 
\end{cases}
\label{eq_hf}
\end{equation}
The reduced coupling of Higgs $H_i$ to the down-type quark $q_d$ ($C_{H_iq_dq_d}$) can be given by
\begin{equation}
    C_{H_iq_dq_d} = 
\begin{cases}
c_{\alpha_1}c_{\alpha_2}/c_{\beta},& i=1, \\
-(c_{\alpha_1}s_{\alpha_2}s_{\alpha_3}+c_{\alpha_3}s_{\alpha_1})/c_{\beta} ,& i=2,  \\
-(c_{\alpha_1}s_{\alpha_2}c_{\alpha_3}-s_{\alpha_1}s_{\alpha_3})/c_{\beta} ,& i=3. 
\end{cases}
\label{eq_hd}
\end{equation}
In this scenario, the couplings of $H_i$ to $b$ quarks and to $\tau$ leptons exhibit distinct dependences on the model parameters, which can naturally account for the observed excesses in the $b\bar{b}$ and $\tau^+\tau^-$ channels. 
These coupling patterns illustrate how the N2HDM-F permits independent control of quark and lepton Yukawa couplings, thereby providing the flexibility required to simultaneously accommodate both $b\bar{b}$ and $\tau^+\tau^-$ excesses.

If $H_1$ is assumed to be the candidate for the possible experimental excesses and $H_2$ is assumed to be the SM-like Higgs with a mass of 125.09~GeV, one can reparameterize the parameters in terms of effective couplings and mixing matrix elements:
\begin{equation}
\begin{aligned}
  m_{H_1},\ m_{H_2},\ m_{H_3},\ m_{H_A},\ m_{H^{\pm}},\ m_{12}^2,\ v_S\\
  C^2_{H_2VV},\ C^2_{H_2t\bar{t}},\ \mathrm{sign}(R_{23}),\ R_{13},\ \tanb,
\end{aligned}
\end{equation}
where $C^2_{H_2VV}$ and $C^2_{H_2t\bar{t}}$ are the effective couplings of $H_2$ to massive gauge bosons and top-quark, $R_{13,23}$ are the mixing matrix elements between $H_{1,2}$ and the singlet field.  

\subsection{Parameter Scan and Constraints}
We performed a random scan over the parameter space of the N2HDM-F using the package \textsf{ScannerS\_v2.0.0}~\cite{Coimbra:2013qq, Muhlleitner:2020wwk}. 
With the mass of $m_{H_2}$ fixed at 125.09 GeV, the remaining 11 parameters were varied within the following ranges:
\begin{align}
  95 &< m_{H_1} < 96~\text{GeV},     && \text{sign}(R_{23}) = \pm1, \nonumber\\
  300 &< m_{H_{3},A} < 1500~\text{GeV}, & 580 &< m_{H^\pm} < 1500~\text{GeV}, \nonumber\\
  0.8 &< \tan\beta < 10,            & -1 &< R_{13} < 1, \label{eq:scan}\\[2pt]
  0.7 &< C_{H_2VV}^2 < 1.0,       & 0.7 &< C_{H_2tt}^2 < 1.2, \nonumber\\
  10^{-3} &< m_{12}^2 < 5\times10^{5}~\text{GeV}^2, & 1 &< v_S < 3000~\text{GeV}. \nonumber
\end{align}
The following constraints are considered:
\begin{itemize}
    \item Constraints from perturbative unitarity: The constraints require that the eigenvalues $\mathcal{M}^i_{2\to2}$ of the $2\to 2$ scattering matrix $\mathcal{M}_{2\to2}$ satisfy
    \begin{align}
        |\mathcal{M}^i_{2\to2}|<8\pi 
    \end{align}
    at tree level~\cite{Kanemura:1993hm}.

    \item Constraints from vacuum stability: The package \textsf{EVADE}~\cite{Hollik:2018wrr, Ferreira:2019iqb} is used to test the stability at tree level.

    \item Constraints from B physics and electroweak precision observables: The global fit result of Ref.~\cite{Haller:2018nnx} on oblique parameters and flavour constraints at 2$\sigma$ level are used.

    \item Constraints from Higgs searches and SM-like Higgs measurements: 
    The \textsf{HiggsBounds-5.10.0} package \cite{Bechtle:2020pkv} and \textsf{HiggsSignal-2.6.0} package \cite{Bechtle:2020uwn} are used to employ the direct searches for additional Higgs and SM-Higgs measurements constraints.
     
\end{itemize}

\subsection{Features of the N2HDM-F Scenario}

\begin{figure*}[tbh] 
\centering 
%\vspace{-20pt} 
\includegraphics[width=\linewidth]{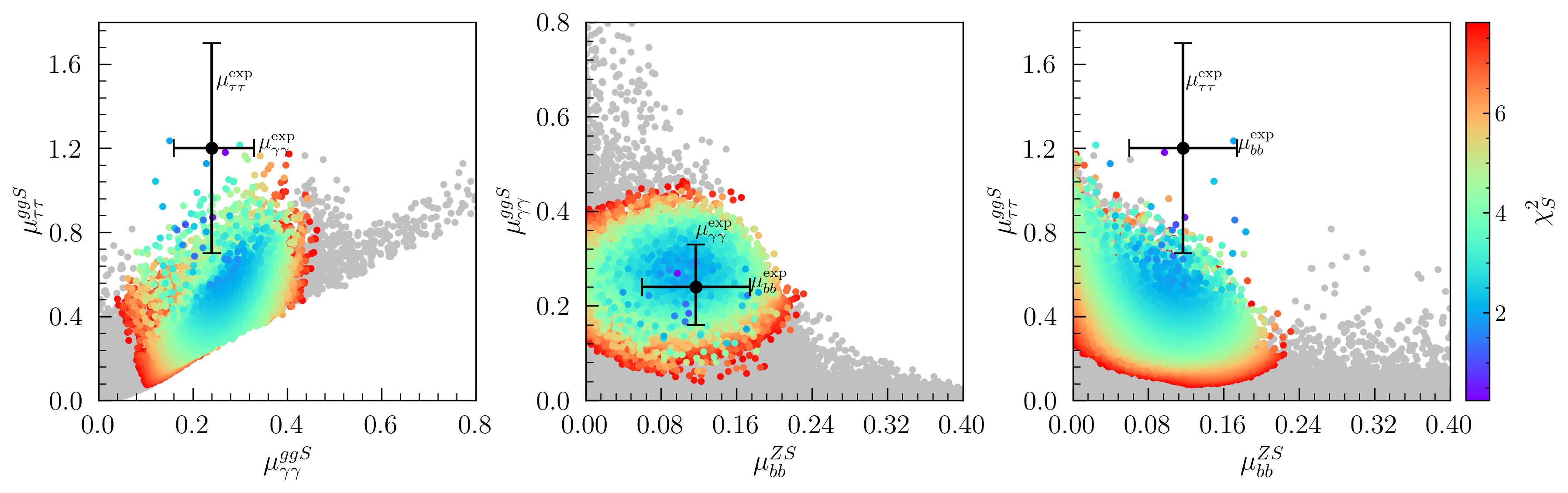}
%\vspace{-35pt} 
\caption{\label{excess}
surviving samples under the above constraints in the $\mu^{ggS}_{\tau\tau}$ versus $\mu^{ggS}_{\gamma\gamma}$ plane (left panel), the $\mu^{ggS}_{\gamma\gamma}$ versus $\mu^{ZS}_{bb}$ plane (middle panel), and $\mu^{ggS}_{\tau\tau}$ versus $\mu^{ZS}_{bb}$ plane (right panel) with colors indicating the fit result of $\chi^2_S$ and the gray samples having $\chi^2_S>7.82$. 
}
\end{figure*}
Fig.~\ref{excess} shows the surviving samples under the above constraints in the $\mu^{ggS}_{\tau\tau}$ versus $\mu^{ggS}_{\gamma\gamma}$ plane (left panel), the $\mu^{ggS}_{\gamma\gamma}$ versus $\mu^{ZS}_{bb}$ plane (middle panel), and $\mu^{ggS}_{\tau\tau}$ versus $\mu^{ZS}_{bb}$ plane (right panel), respectively. 
The colors indicate the fit results of $\chi^2_S$ \footnote{In constructing $\chi^2_S$, we treat the three excesses as independent Gaussian measurements and neglect possible correlations among the experimental systematics. 
This is sufficient for our purpose of identifying representative benchmark scenarios and should not be interpreted as a rigorous global statistical combination.}, defined as:
\begin{align}
    \chi^2_S = 
    \frac{(\mu_{\tau\tau}^{ggS} -\mu_{\tau\tau}^{\mathrm{exp}})^2}{(\Delta \mu_{\tau\tau}^{\mathrm{exp}})^2} + 
    \frac{(\mu_{\gamma\gamma}^{ggS} -\mu_{\gamma\gamma}^{\mathrm{exp}})^2}{(\Delta \mu_{\gamma\gamma}^{\mathrm{exp}})^2} + 
    \frac{(\mu_{bb}^{ZS} -\mu_{bb}^{\mathrm{exp}})^2}{(\Delta \mu_{bb}^{\mathrm{exp}})^2},
\end{align}
where $\mu_{xx}^{\mathrm{exp}}$ and $\Delta\mu_{xx}^{\mathrm{exp}}$ are the central values and their corresponding $1\sigma$ uncertainties for the experimental excesses in the $\tau^+\tau^-$, $b\bar{b}$, and $\gamma\gamma$ channels, respectively. 
These are parameterized as $\mu_{\tau\tau}^{\mathrm{exp}}=1.2_{-0.5}^{+0.5}$ \cite{CMS:2022goy}, $\mu_{bb}^{\mathrm{exp}}=0.117\pm0.057$ \cite{LEPWorkingGroupforHiggsbosonsearches:2003ing}, and $\mu_{\gamma\gamma}^{\mathrm{exp}}=0.24^{+0.09}_{-0.08}$ \cite{CMS:2024yhz, ATLAS:2024bjr}, and are marked in the Fig.~\ref{excess}.
The $\mu_{xx}^{ggS}$ and $\mu_{xx}^{ZS}$ are the signal strength in the gluon fusion channel ($gg\to S\to xx$) and Higgsstrahlung channel ($ee\to ZS(\to xx)$) and can be parameterized as
\begin{align}
    \mu_{xx}^{ggS} = \frac{\sigma(gg\to S) \times \mathrm{Br}(S\to xx)}{\sigma_{\mathrm{SM}}(gg\to h_{95}) \times \mathrm{Br_{SM}}(h_{95}\to xx)}
\end{align}
and
\begin{align}
    \mu_{xx}^{ZS} = \frac{\sigma(e^+e^-\to ZS) \times \mathrm{Br}(S\to xx)}{\sigma_{\mathrm{SM}}(e^+e^-\to Zh_{95}) \times \mathrm{Br_{SM}}(h_{95}\to xx)},
\end{align}
where $\sigma$ and $\mathrm{Br}$ denote the production cross section and branching ratio in the Beyond the Standard Model (BSM) scenario, respectively, while $\sigma_{\mathrm{SM}}$ and $\mathrm{Br}_{\mathrm{SM}}$ represent the corresponding SM values.
All three excesses can be elegantly explained within the broken phase N2HDM-F scenario. The best-fit point has a minimum $\chi^2_S$ value of 0.24\,.

\begin{figure}[!tbh] 
\centering 
%\vspace{-1pt} 
\includegraphics[width=0.7\linewidth]{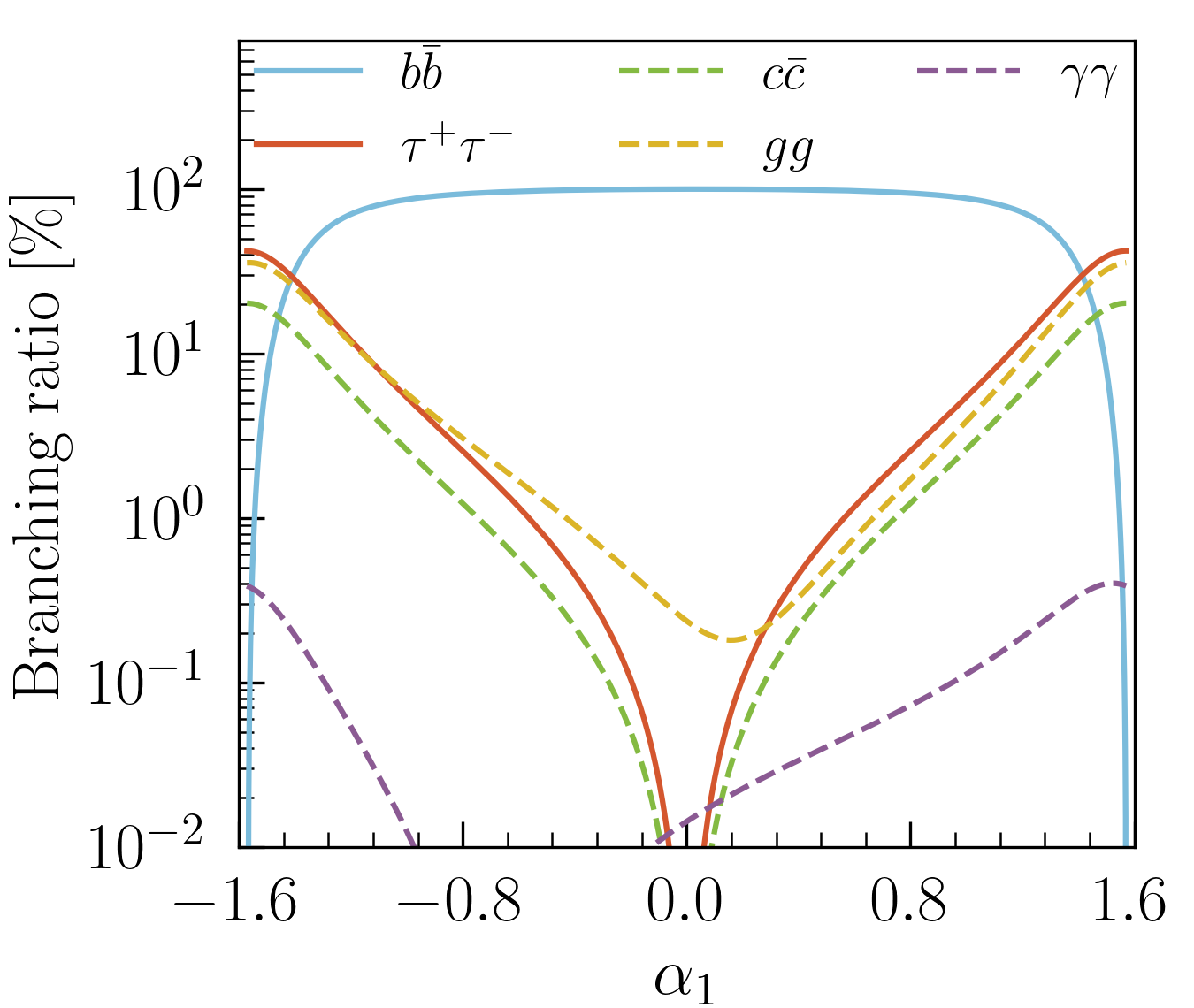}
%\vspace{-10pt} 
\caption{\label{Br}
The branching ratios of the light Higgs $S$ to $b\bar{b}$, $c\bar{c}$, $\gamma\gamma$, $\tau^+\tau^-$, and $gg$ in the N2HDM-F as a function of $\alpha_1$, for the benchmark points with $\cos\alpha_2 = \sqrt{2}/2$ and $\tan\beta = 2$.
}
\end{figure}

Fig.~\ref{Br} shows the branching ratios of the light Higgs $S$ to $b\bar{b}$, $c\bar{c}$, $\gamma\gamma$, $\tau^+\tau^-$, and $gg$ as a function of $\alpha_1$ in the N2HDM-F, for benchmark points with $\cos\alpha_2 = \sqrt{2}/2$ and $\tan\beta = 2$.
These branching ratios are computed using the package \textsf{N2HDECY}~\cite{Engeln:2018mbg}. 
It can be seen that $S$ decays predominantly to $b\bar{b}$ over most of the $\alpha_1$ range, except near $\alpha_1 \approx \pm\pi /2$ (where $\cos\alpha_1 \approx 0$). 
In this region, the dominant decay channels become $\tau^+\tau^-$, $c\bar{c}$, and $gg$.
However, the $c\bar{c}$ and $gg$ channels are associated with a larger SM background from $Z$ boson hadronic decays~\cite{ParticleDataGroup:2024cfk}. 
Considering the channels where experimental signal excesses have been observed, we focus on the $b\bar{b}$ and $\tau^+\tau^-$ channels in the subsequent analysis.
The combination of these two channels ensures good coverage across the entire $\alpha_1$ parameter space.

\section{\label{sec:mc}Monte Carlo Simulation} 
\subsection{Signal and Background Modeling}
We focus on the Higgsstrahlung process $e^+e^- \to Z(\to\mu^+\mu^-)S(\to\tau^+\tau^-(b\bar{b}))$ at the CEPC with a center-of-mass energy of 240 GeV.
In this signal process, the hypothetical scalar $S$ is produced in association with a $Z$ boson, where $S$ subsequently decays into a pair of $\tau$ leptons or $b$ quarks, and the $Z$ boson decays into a muon pair.
Although the dominant decay channel of the $Z$ boson is hadronic and invisible \cite{ParticleDataGroup:2024cfk}, we instead select the dimuon final state to utilize the recoil mass technique for reconstructing the mass of $S$.
The recoil mass $M_{\mathrm{recoil}}$ is defined as ~\cite{Chen:2016zpw}
\begin{align}\label{recoil_m}
    M_{\mathrm{recoil}} \equiv \sqrt{s + M_{\mu^+\mu^-}^2 - 2\sqrt{s}(E_{\mu^+} + E_{\mu^-})},
\end{align}
where $\sqrt{s}$ is the collision energy, $M_{\mu^+\mu^-}^2$ is the square of invariant mass of dimuon, and $E_{\mu}$ is the energy of muon. 
Other useful observables can be found in Ref.~\cite{Yang:2022qga}.
One can also choose $Z$ boson decays into a pair of electrons to calculate the recoil mass, and this situation will be discussed later in the paper.

\begin{figure*}[!tbp] 
\centering 
%\vspace{-20pt} 
\includegraphics[width=1\linewidth]{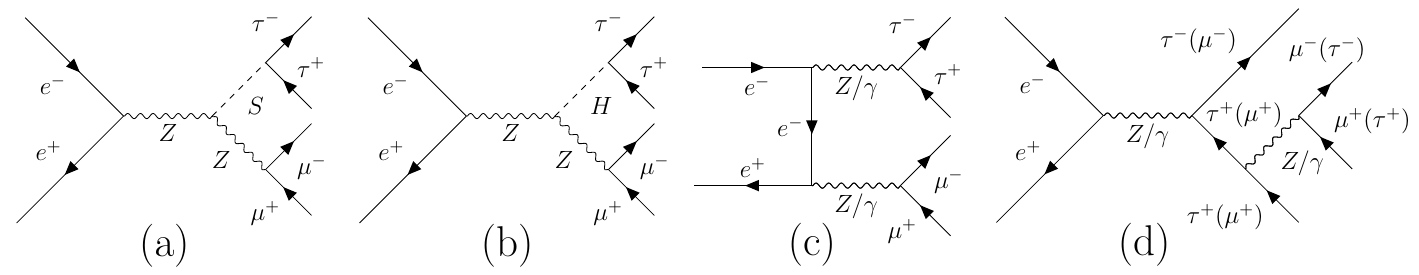}
%\vspace{-35pt} 
\caption{
Representative Feynman diagrams of the signal and irreducible backgrounds in the $\tau^+\tau^-$ decay channel, where panel (a) corresponds to the signal, and panels (b), (c), and (d) represent the irreducible backgrounds.
Diagrams in panels (c) and (d) also imply those related by fermion–antifermion exchange, 
$\mu \leftrightarrow \tau$ interchange, and $Z \leftrightarrow \gamma$ exchange.
}\label{feyn}
\end{figure*}

For the $\tau^+\tau^-$ decay channel, the dominant irreducible SM backgrounds originate from the process $e^+ e^- \to \mu^+ \mu^- \tau^+ \tau^-$, in which the $\mu^+\mu^-$ and $\tau^+\tau^-$ pairs can decay from a $Z$ boson or a virtual photon. 
Additionally, the $\tau^+\tau^-$ pair may also come from the decay of the SM-like Higgs boson. 
The leading Feynman diagrams for the signal and the irreducible backgrounds in the $\tau^+\tau^-$ channel are shown in Fig.~\ref{feyn}, where panel (a) corresponds to the signal, and panels (b), (c), and (d) represent the irreducible backgrounds.
We also consider the reducible SM background from $e^+e^- \to \mu^+\mu^- xx$, where $x$ denotes a final-state particle from a light-flavor quark, a $b$-quark, an electron, or a muon, which could be misidentified as a tau lepton in the detector.

\begin{figure*}[!tbp] 
\centering 
%\vspace{-10pt} 
\includegraphics[width=\linewidth]{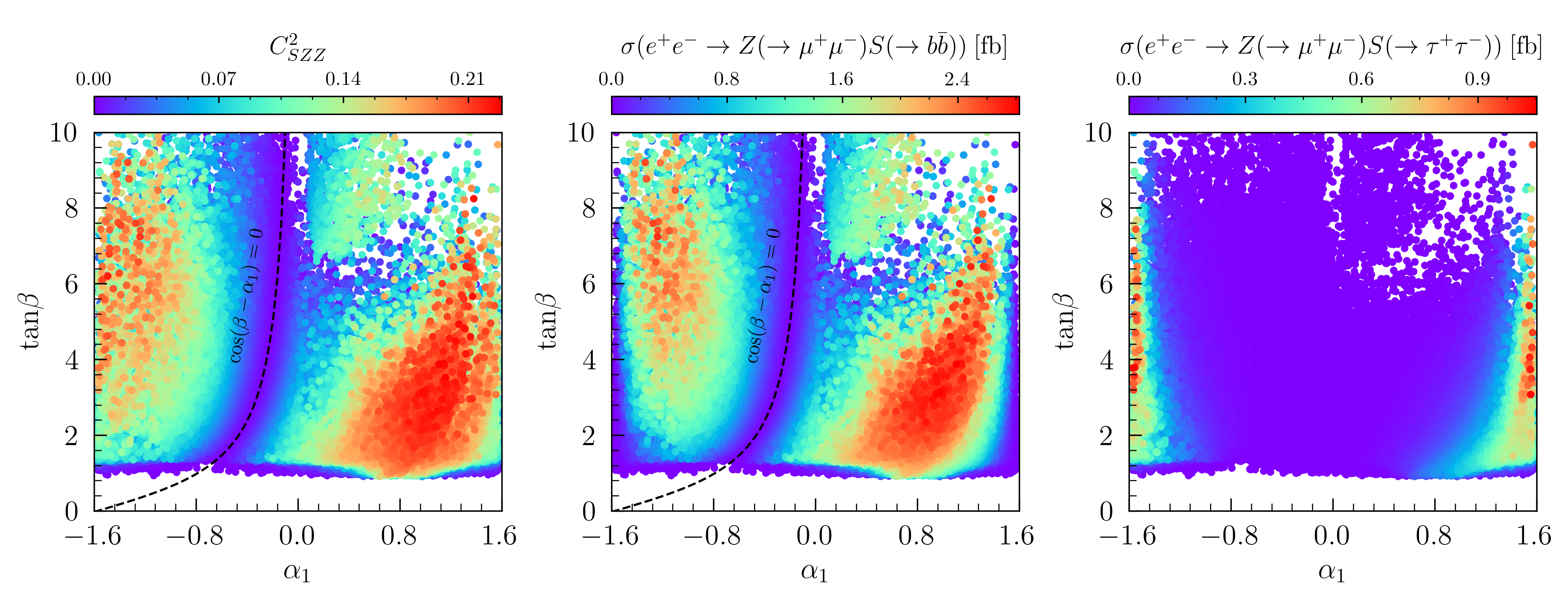}
%\vspace{-35pt} 
\caption{\label{cross_section}
Surviving samples in the tan$\beta$ versus $\alpha _1$ plane with the colors indicating the square of the reduced coupling of $S$ to $Z$ boson (left panel), the cross section of the $b\bar{b}$ channel (middle panel), and the cross section of the $\tau^+\tau^-$ channel (right panel).
}
\end{figure*}

In the case of the $b\bar{b}$ decay channel, the background sources are analogous to those in the $\tau^+\tau^-$ channel, with the roles of $b\bar{b}$ and $\tau^+\tau^-$ are interchanged. 
Therefore, we adopt a common background process for both channels: $e^+e^- \to \mu^+\mu^- xx$, where $x = u, d, c, s, b, e, \mu, \tau$. 
These background processes comprise two components: a resonant part, in which the muon pair comes from a $Z$ boson, and a non-resonant part, where the muon pair originates from a virtual photon.

Fig.~\ref{cross_section} shows the surviving samples in the tan$\beta$ versus $\alpha _1$ planes with the colors indicating the square of the reduced coupling of $S$ to the $Z$ boson $C_{SZZ}^2$ (left panel), the cross section of the $b\bar{b}$ channel (middle panel), and the cross section of the $\tau^+\tau^-$ channel (right panel). 
The cross section for the $b\bar{b}$ and $\tau^+\tau^-$ channels is computed as
\begin{equation}
\begin{aligned}
    \sigma =\ &\sigma_{SM}(e^+e^-\to ZS)|_{m_h=95.5\,\mathrm{GeV}}  \times C_{SZZ}^2 \\
    &\times \mathrm{Br}(Z\to\mu^+\mu^-) \cdot \mathrm{Br}(S\to \tau^+\tau^-(b\bar{b})),
\end{aligned}
\end{equation}
where $\sigma_{\mathrm{SM}}(e^+e^-\to ZS)|_{m_h =\, 95.5\ \mathrm{GeV}}$ is the SM cross section for this process with a Higgs mass of 95.5 GeV, calculated at leading order using the \textsf{MadGraph5\_aMC@NLO\_v3.6.2} package~\cite{Alwall:2011uj, Alwall:2014hca}. The branching ratio of $\mathrm{Br}(Z\to \mu^+\mu^-)$ takes the SM value of 3.3662\% as given by the PDG~\cite{ParticleDataGroup:2024cfk}.  
The $C_{SZZ}^2$ and $\mathrm{Br}(S\to \tau^+\tau^-(b\bar{b}))$ provided by the N2HDM-F.

As shown in the left panel, the squared reduced coupling $C_{SZZ}^2$ can reach up to 0.23 for the samples near $\alpha_1\approx0.8$.
In contrast, $C_{SZZ}^2$ approaches zero for samples with cos$(\beta - \alpha_1) \approx 0$, where the light Higgs $S$ couples weakly to massive gauge bosons and could escape detection at the CEPC.
The middle and right panels indicate that the cross section for the $b\bar{b}$ channel can reach 2.8 fb for the samples with $\alpha_1\approx 0.8$, while that for the  $\tau^+\tau^-$ channel reaches 1.1 fb near $\alpha_1\approx \pm{\pi/2}$.
As anticipated, except for the samples near cos$(\beta - \alpha_1)\approx 0$, where the cross section of the $b\bar{b}$ channel is suppressed, the cross section of the $\tau^+\tau^-$ channel becomes comparatively larger, as can be inferred from a combined interpretation of all three panels. 
In the following MC simulations, the cross sections for the $\tau^+\tau^-$ and $b\bar{b}$ decay channels are assumed to be $1/5$ of their SM values as a benchmark point.

\subsection{Event Generation and Selection in the \texorpdfstring{$\tau^+\tau^-$}{tau-tau} Channel}

The \textsf{MadGraph5\_aMC@NLO\_v3.6.2} package~\cite{Alwall:2011uj, Alwall:2014hca} is used to perform the MC simulation. During the simulation, jet clustering is performed using the \textsf{ee-$k_T$} algorithm through \textsf{FastJet} \cite{Cacciari:2008gp}, and the tau jet tagging efficiency adopted the lower limit $80\%$ as given in Ref.~\cite{Yu:2020bxh}.  
The particle decays are handled by \textsf{PYTHIA\_v8.2} \cite{Sjostrand:2014zea} through the \textsf{MG5a-MC\_PY8\_interface}.  
The CEPC detector response is simulated using the CEPC 4th-card with \textsf{Delphes\_v3.5.0} \cite{deFavereau:2013fsa, Selvaggi:2014mya}.
More than one million events of signal and background are generated through the MC simulation. 

In the event selection, two muons in the final state are required to calculate the recoil mass $M_{\mathrm{recoil}}$ to reconstruct the mass of $S$.
We apply the following $\textbf{Basic Cut}$ to the events:
\begin{equation}\label{basic_cut}
\begin{aligned}
    N(\mu)\geq\ 2,\ |\eta(\mu)| <\ 2.5,\ p_{\mathrm{T}}({\mu}) >\ 10\  \mathrm{GeV}, \\
    \Delta R(\mu_1,\mu_2) >\ 0.4,\ 66<M_{\mu\mu}<116\ \mathrm{GeV},
\end{aligned}
\end{equation}
where $\eta$, $\phi$, and $p_\mathrm{T}$ denote the pseudorapidity, azimuthal angle, and transverse momentum, respectively; $\Delta R(\mu_1,\mu_2) = \sqrt{(\eta_{\mu_1}-\eta_{\mu_2})^2 + (\phi_{\mu_1}+\phi_{\mu_2})^2}$ gives the angular separation between the two muons; $M_{\mu\mu}=\sqrt{(E_{\mu_1} + E_{\mu_2})^2-(\vec{p}_{\mu_1}+\vec{p}_{\mu_2})^2}$ is the invariant mass of the muon pair.

In the signal process, the muon pair originates from the $Z$ boson decay, and its invariant mass distribution is expected to peak near the $Z$ boson mass.
In contrast, a significant fraction of the background events involve muon pairs not produced from a $Z$ boson resonance, and thus do not exhibit such a peak in their invariant mass distribution.
To suppress this non-resonant background component and ensure robust signal retention, we require that $66<M_{\mu\mu}<116\ \mathrm{GeV.}$

\begin{figure*}[!tbp] 
\centering 
\includegraphics[width=0.7\linewidth]{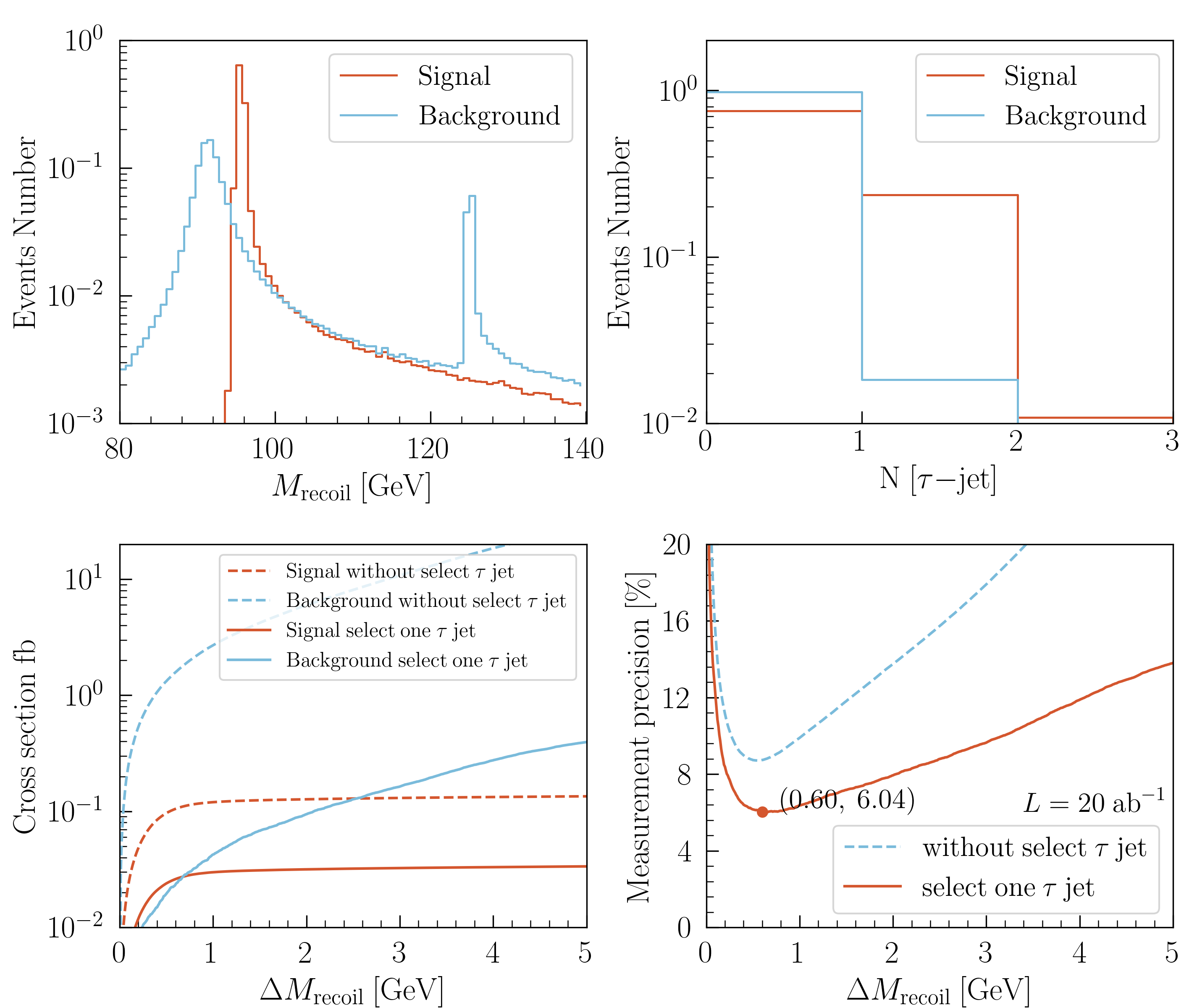}
\caption{
Normalized event distributions for the signal and background in the recoil mass $M_{\mathrm{recoil}}$ (upper left panel) and the $\tau$-jet number N [$\tau$-jet] (upper right panel), along with the signal and background cross sections under different selections on these variables (lower left panel), as well as the corresponding measurement precision of the signal strength after applying these selections (lower right panel) in the $\tau^+\tau^-$ channel.
}\label{rec_d}
\end{figure*}

The normalized event distributions of the recoil mass ($M_{\mathrm{recoil}}$) and the number of $\tau$-jet (N [$\tau$-jet]) for signal and background after the Basic cut are shown in the upper left and upper right panels of Fig.~\ref{rec_d}, respectively. 
The upper left panel shows that the signal $M_{\mathrm{recoil}}$ distribution peaks sharply around 95.5 GeV, consistent with the mass of the light Higgs $S$. 
In contrast, the background distribution exhibits peaks near 91 GeV and 125 GeV, corresponding to the masses of the $Z$ boson and the SM-like Higgs boson, respectively, with the peak at 91 GeV being considerably broader.
The upper right panel indicates that only about 2\%  of background events contain at least one $\tau$-jet, whereas for the signal, this fraction reaches approximately 25\%. 
Therefore, applying selections on $M_{\mathrm{recoil}}$ and the number of $\tau$-jets can effectively suppress the background.

A fine-grained scan of the $\Delta M_{\mathrm{recoil}}$ window from 0 to 5 GeV is performed. The signal region is defined as $|M_{\mathrm{recoil}} - 95.5\ \mathrm{GeV}| < \Delta M_{\mathrm{recoil}}$, and the cross sections of signal and background are calculated based on the number of surviving events in each window.
The lower-left panel of Fig.~\ref{rec_d} shows the signal and background cross sections after applying different selections on the recoil-mass window $\Delta M_{\rm recoil}$ and on the number of $\tau$-jets (N[$\tau$-jet]).
The corresponding statistical precision of the signal strength measurement is shown in the lower right panel, calculated using \cite{Zheng:2020ult}:
\begin{align}\label{prec}
    P =\frac{\sigma(\mu)}{\mu} = \frac{\sqrt{S+B}}{S},
\end{align}
where $S$ and $B$ denote the number of expected signal and background events for different $\Delta M_{\mathrm{recoil}}$ windows, and both are calculated by their cross sections, assuming an integrated luminosity of 20~ab$^{-1}$ at the CEPC.

As shown in the lower left panel of Fig.~\ref{rec_d}, the background cross section exceeds that of the signal by more than one order of magnitude if no selection is applied on the number of $\tau$-jets, regardless of the $M_{\mathrm{recoil}}$ cut. 
However, by selecting events with at least one $\tau$-jet and applying an appropriate $M_{\mathrm{recoil}}$ cut, the signal and background cross sections can be brought to the same order of magnitude. 
Furthermore, for any given $M_{\mathrm{recoil}}$ cut, the measurement precision is significantly improved when requiring a $\tau$-jet in the final state. 
The optimal precision of 6.04\% is achieved for events containing a $\tau$-jet and satisfying $\Delta M_{\mathrm{recoil}} \leq 0.60$ GeV. The choice reflects a balance between background suppression and signal retention: while a narrower window reduces background more effectively, it also excludes a significant fraction of signal events, thereby degrading the overall precision. Based on this, we apply the following $\bm \tau$\textbf{-jet cut} and \textbf{$\bm{M_{\mathrm{recoil}}}$ Cut} that require

\begin{align}
    \mathrm{N} [\tau\mathrm{-jet}] \geq 1
\end{align}
and 
\begin{align}
    \Delta M_{\mathrm{recoil}} = 0.60\ \mathrm{GeV}.
\end{align}
Table~\ref{t01} shows the cut flow for the $\tau^+\tau^-$ decay channel of the signal and background at the 240 GeV CEPC with an integrated luminosity of $L=20\abm$.
After the $\bm \tau$\textbf{-jet cut}, the background is suppressed by approximately two orders of magnitude while about 25\% of the signal is retained, improving the signal strength measurement precision from 32.56\% to 18.41\%. A subsequent \textbf{$\bm{M_{\mathrm{recoil}}}$ Cut} further suppresses the background by over an order of magnitude and preserves about 67\% of the signal, leading to a final precision of 6.04\%.

\begin{table}[!tbp]
\centering
\caption{Cut flow of signal and background in the $\tau^+\tau^-$ channel at 240 GeV CEPC with $L=20\abm$}
\label{t01}
\begin{tabular*}{\columnwidth}{@{\extracolsep{\fill}}ccccccccc}
    \toprule
    \multirow{2}{*}{Cuts} & \multicolumn{2}{c}{Cross section [fb]} & \multirow{2}{*}{Precision} \\ 
    \cmidrule(r){2-3} 
    & Signal & Background & \\ \midrule
    Initial & 0.218 & 84.250 & 29.81$\%$ \\ 
    Basic & 0.159 & 53.373 & 32.56$\%$ \\ 
    $\tau$-$jet$ & 0.039 & 1.000 & 18.41$\%$ \\ 
    $M_{\mathrm{recoil}}$ & 0.026 & 0.024 & 6.04$\%$ \\  
    \bottomrule
\end{tabular*}
\end{table}

\begin{figure*}[!tbp] 
\centering 
\includegraphics[width=1\linewidth]{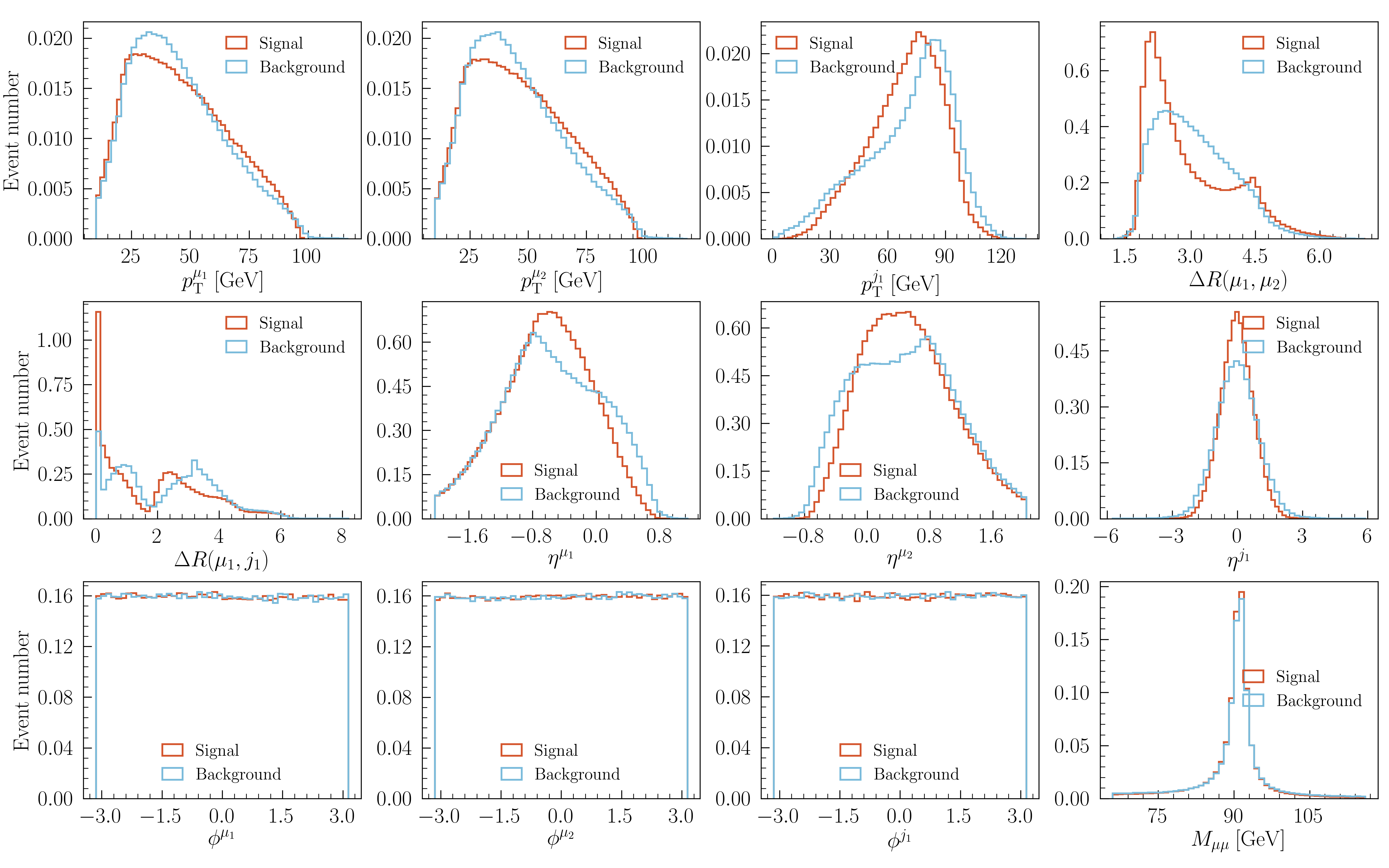}
\caption{\label{dis}
The normalized distributions of the transverse momenta, pseudorapidities, and azimuthal angles of the two muons and the leading jet, as well as the angular separations $\Delta R$ between the two muons and between the leading muon and the leading jet, after applying the Basic cut.
}
\end{figure*}

After applying the Basic cut, Fig.~\ref{dis} shows the normalized distributions of the transverse momenta, pseudorapidities, and azimuthal angles of the two muons and the leading jet, as well as the angular separations $\Delta R$ between the two muons and between the leading muon and the leading jet. 
As shown in this figure, the significant overlap between the signal and background means that further cuts on these kinematic observables do not lead to a significant improvement in signal significance or measurement precision.

\begin{figure*}[!tbh] 
\centering 
%\vspace{-5pt} 
\includegraphics[width=0.7\linewidth]{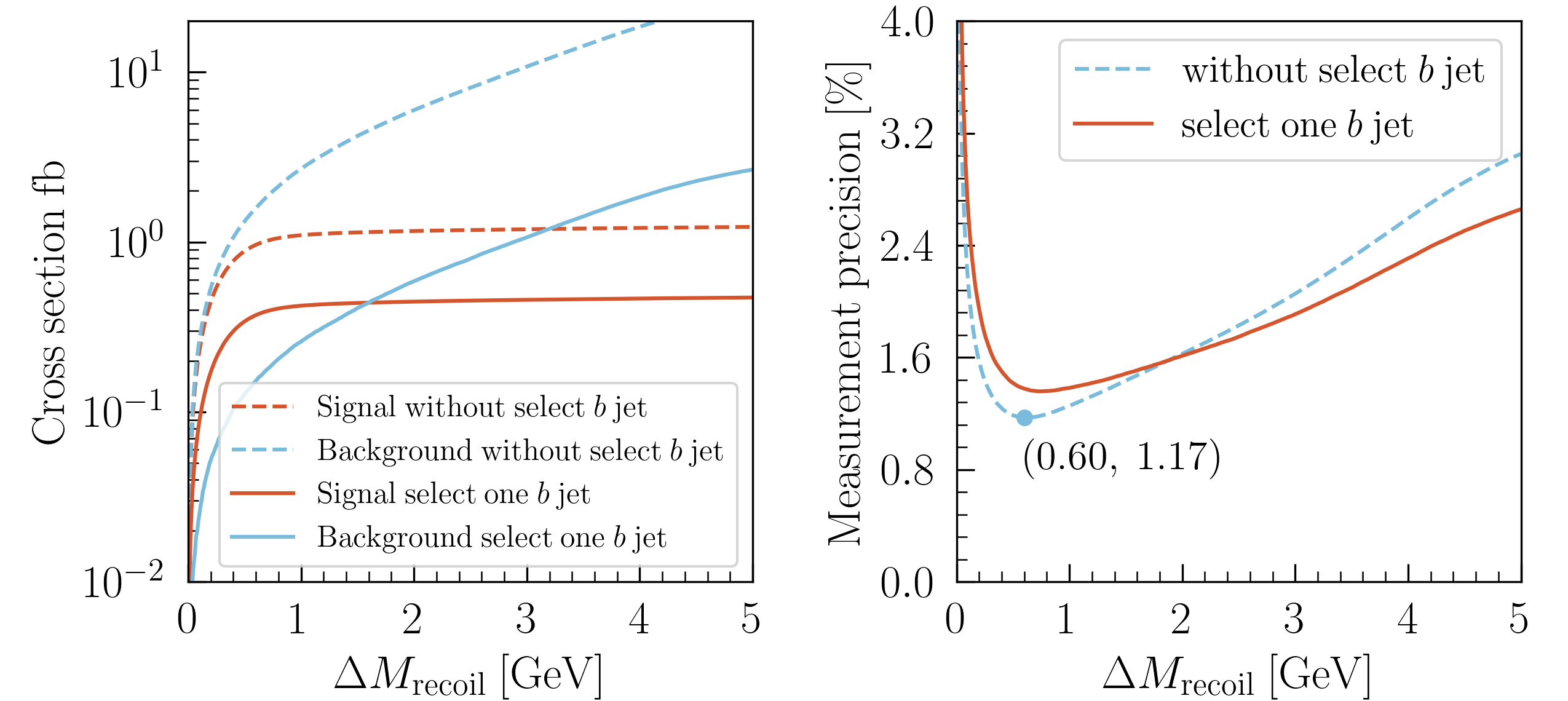}
%\vspace{-15pt} 
\caption{\label{bb_rec}
The cross sections of the signal and background (left panel), and the signal strength measurement precision (right panel) under different selections on the $M_{\mathrm{recoil}}$ and N [$b$-jet] in the $b\bar{b}$ channel.
}
\end{figure*}

\subsection{Event Generation and Selection in the \texorpdfstring{$b\bar{b}$}{bb} Channel}
The event generation for the $b\bar{b}$ channel follows the same procedure as that for the $\tau^+\tau^-$ channel, with the $b$-jet tagging efficiency also set to 80\%. In the event selection, the $\textbf{Basic Cut}$ is also the same as that in the $\tau^+\tau^-$ channel. 
The only difference is that selecting a $b$-jet in the final state can reduce the reducible background.
Fig.~\ref{bb_rec} shows the cross sections for signal and background processes (left panel) and the corresponding precision on the signal strength (right panel) after applying different selections on the recoil-mass window $\Delta M_{\rm recoil}$ and on the number of $b$-jets (N[$b$-jet]).
Unlike the $\tau^+\tau^-$ decay channel, the signal and background cross sections become comparable by applying only the $M_{\mathrm{recoil}}$ selection.
Selecting events that contain a $b$-jet suppresses the background by approximately one order of magnitude, but also reduces the signal efficiency. 
Consequently, this combined selection fails to achieve the maximum measurement precision attainable with the $M_{\mathrm{recoil}}$ selection alone.
The signal strength can be measured with a precision of 1.17\% by selecting events with $\Delta M_{\mathrm{recoil}} = $ 0.60 GeV and without requiring a $b$-jet in the final state. This choice of window is optimal to maximize the measurement precision because a narrower window would discard too much signal, while a wider one would introduce excessive background.
Then the \textbf{$\bm{M_{\mathrm{recoil}}}$ Cut} are performed that require the event be satisfied for 
\begin{align}
    |M_{\mathrm{recoil}}-95.5| \leq 0.60\ \mathrm{GeV}.
\end{align}

The cut flow for the $b\bar{b}$ decay channel of the signal and background at the 240 GeV CEPC with an integrated luminosity of $L=20\abm$ is shown in Table~\ref{t02}. It can be seen that the \textbf{$\bm{M_{\mathrm{recoil}}}$ Cut} suppresses the background more than one order of magnitude while retaining about 69\% of the signal. The corresponding measurement precision is improved from 3.76\% to 1.17\%.
The $b\bar{b}$ channel is also plagued by a limitation similar to that in the $\tau^+\tau^-$ channel. 
Despite successive cuts on jet-level kinematic observables, the measurement precision does not improve significantly.

\begin{table}[!tbp]
\centering
\caption{Cut flow of signal and background in the $b\bar{b}$ channel at 240 GeV CEPC with $L=20\abm$}
\label{t02}
\begin{tabular*}{\columnwidth}{@{\extracolsep{\fill}}lccc}
    \toprule
    \multirow{2}{*}{Cuts} & \multicolumn{2}{c}{Cross section [fb]} & \multirow{2}{*}{Precision} \\ 
    \cmidrule(r){2-3} 
    & Signal & Background & \\ \midrule
    Initial & 2.099 & 84.250 & 3.13\% \\ 
    Basic & 1.391 & 53.373 & 3.76\% \\ 
    % $M_{\mu\mu}$ & 1.454 & 59.929 & 3.81\% \\ 
    $M_{\mathrm{recoil}}$ & 0.963 & 1.585 & 1.17\% \\ 
    \bottomrule
\end{tabular*}
\end{table}

In fact, through a systematic exploration of additional selection criteria for $\tau^+\tau^-$ and $b\bar{b}$ channels, we find that imposing additional cuts on fundamental jet- or lepton-level observables does not translate into a noticeable improvement in precision over the presented results. This provides strong evidence that the current cut-based analysis is close to optimally configured.
The further improvement of measurement precision using the cut-based method in the $\tau^+\tau^-$ and $(b\bar{b})$ decay channels is primarily constrained by two technical limitations:
\begin{itemize}
    \item First, despite a general $\tau$-tagging efficiency of up to 80\%, only about 25\% of hadronically decaying $\tau$ leptons are successfully identified as $\tau$-jets in the $\tau^+\tau^-$ channel, leading to significant signal loss. While the situation in the $b\bar{b}$ channel is better, it similarly suffers from a substantial loss of signal events. 
    
    \item Second, event selection that depends on reconstructed jet information inevitably discards a considerable amount of information from the final-state particles, some of which may be critical. Furthermore, the high multiplicity and diversity of final-state particles make it challenging for the cut-based method to fully exploit all available information and capture the underlying correlations among them. 
\end{itemize}

To enhance the search sensitivity for the 95 GeV light Higgs boson and improve the measurement precision, we employ deep neural networks (DNNs), specifically the ParT \cite{Qu:2022mxj} and MIParT \cite{Wu:2024thh} architectures.
Both architectures are based on the Transformer model, which excels at capturing complex correlations within the data.

\section{\label{sec:MIParT}Application of Transformer Networks} 

ParT~\cite{Qu:2022mxj} incorporates pairwise particle interactions into the attention mechanism, achieving superior particle tagging performance compared to standard Transformer architectures and previous ML models. 
Building upon ParT, MIParT~\cite{Wu:2024thh} introduces a More-Interaction Attention mechanism that increases the dimensionality of particle interaction embeddings. 
This enhancement not only significantly reduces the model size but also leads to further performance gains.
These methods can exploit a richer and more complex set of information from the event data compared to cut-based analysis.

For the cut-based analysis, the available information comprises jet-level observables (e.g., jet four-momenta), lepton-level observables (e.g., muon four-momenta), and event-level variables (e.g., the invariant mass and recoil mass of the dimuon system). 
ParT and MIParT can both train directly on final-state particle-level features.

In our ML-based analysis, events are first preselected according to the $\textbf{Basic Cut}$ of Eq.\ref{basic_cut}, and the ParT and MIParT networks are subsequently trained on these events using the following event-level features:
\begin{itemize}
    \item The recoil mass of two muons: $M_{\mathrm{recoil}}$.
    \item The invariant mass of two muons: $M_{\mu\mu}$.
    \item The average azimuthal angle of two muons \footnote{ The reason why we use averaged quantities such as $\phi = (\phi_1 + \phi_2)/2$, is that the architecture of the original ParT and MIParT networks needs a reference direction. 
    We therefore define the reference direction as the average of the two muon directions. 
    We have tested that the choice of reference direction has almost no impact on the network's performance.}: $\phi = (\phi_{\mu_1} + \phi_{\mu_2})/2$.
    \item The average pseudorapidity of two muons: $\eta = (\eta_{\mu_1} + \eta_{\mu_2})/2$.
\end{itemize}

The following features at the particle-level are also used:
\begin{itemize}
    \item The transverse momentum of the final state particles: $P_T^{\mathrm{part}}$.
    \item The energy of the final state particles: $E^{\mathrm{part}}$.
    \item The renormalized transverse momentum of the final state particles: $R_{P_T^{\mathrm{part}}}=P_T^{\mathrm{part}}/(P_T^{\mu_1} + P_T^{\mu_2})$.
    \item The renormalized energy of the final state particles: $R_{E^{\mathrm{part}}}=E^{\mathrm{part}}/(E^{\mu_1} + E^{\mu_2})$.
    \item The difference between the azimuthal angle of the final state particles and the average azimuthal angle of two muons $\phi$: $\Delta \phi^{\mathrm{part}}=\phi^{\mathrm{part}}-\phi$. 
    \item The difference between the pseudorapidity angle of the final state particles and the average pseudorapidity of two muons $\eta$: $\Delta \eta^{\mathrm{part}}=\eta^{\mathrm{part}}-\eta$.
    \item The angular separations between the final state particles to the event $\eta$ and $\phi$: $\Delta R^{\mathrm{part}}=\sqrt{(\eta^{\mathrm{part}}-\eta)^2+(\phi^{\mathrm{part}}-\phi)^2}$. 
    \item The charge of the final state particles: $C^{\mathrm{part}}=\pm 1$ for charged particle and $C^{\mathrm{part}}= 0$ for neutral particle.
    \item The particle identification of the final-state particles: charged hadron, neutral hadron, photon, muon, or electron. 
\end{itemize}

\begin{figure*}[!tbp] 
\centering 
\includegraphics[width=0.9\linewidth]{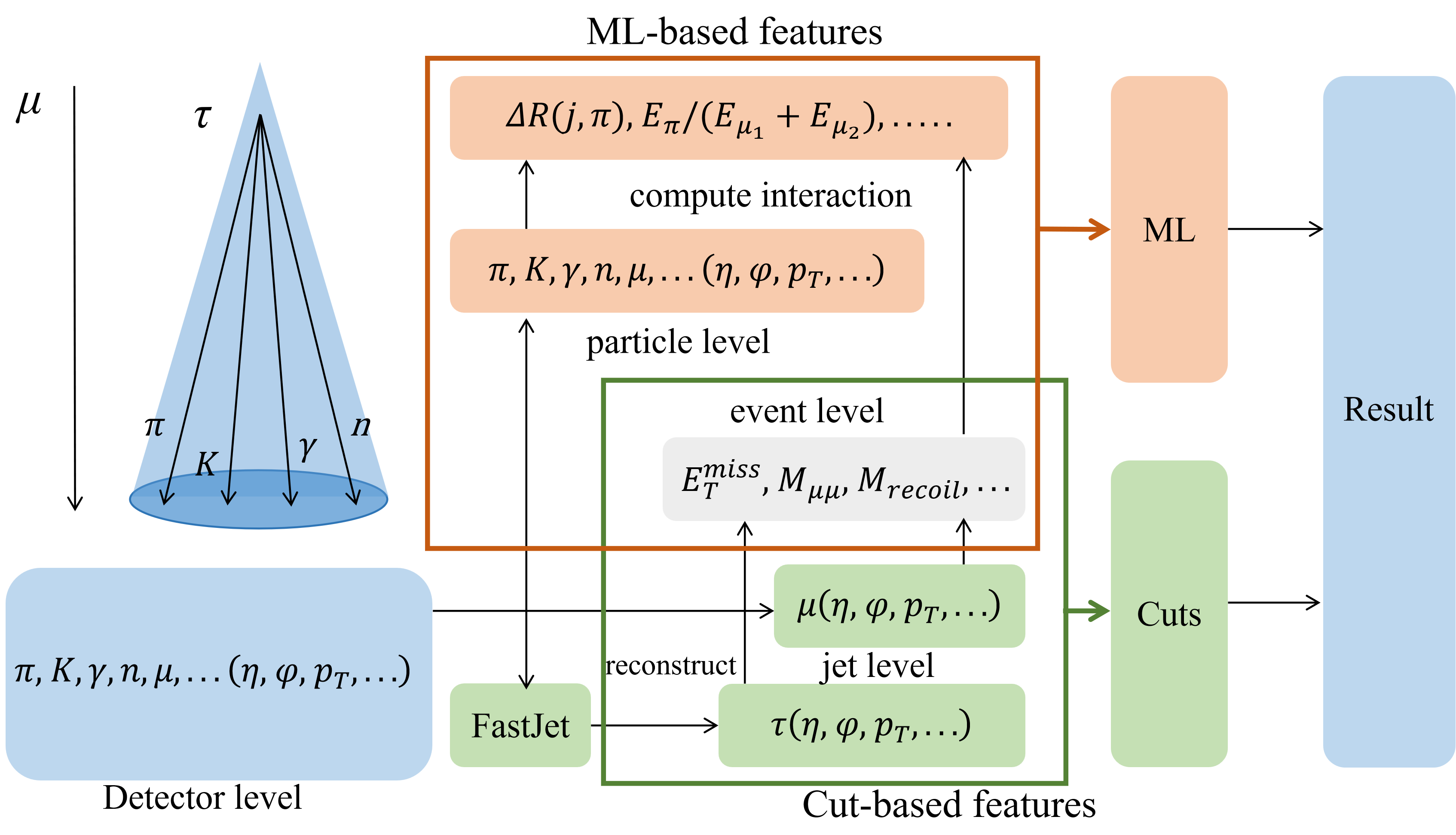}
\caption{\label{process}
Schematic comparison of the available features used in jet-level cut-based and particle-level ML analysis.
}

\end{figure*}

A schematic diagram comparing the available features used in the jet-level cut-based analysis versus our particle-level ML-based analysis is presented in Fig.~\ref{process}.
Compared to the cut-based analysis, which only utilizes information from jets, muons, and event-level variables, our ML-based analysis incorporates a wider set of features at the particle-level. 
These features are difficult to leverage effectively with cut-based methods but can be fully exploited by models such as ParT and MIParT, which possess strong information processing capabilities.

During data preprocessing, the following features were standardized after applying a logarithmic transformation: $p_T^{\mathrm{part}}$, $E^{\mathrm{part}}$, $R_{p_T^{\mathrm{part}}}$, and $R_{E^{\mathrm{part}}}$. To reduce computational cost, the maximum number of particles per event was set to 50. 
We constructed a balanced dataset of over 300,000 events (with 150,000 signal and 150,000 background events) for training and evaluating the ParT and MIParT models. 
The dataset was split into a training set (90\% of the data) and a validation set (10\%). 
Other architectural hyperparameters, such as the embedding dimension, number of transformer blocks, number of attention heads, and hidden dimension in the feed-forward network, followed the default configurations used in ParT \cite{Qu:2022mxj} and MIParT \cite{Wu:2024thh}. 
The models were trained with a batch size of 256 and an initial learning rate of 0.001. 
After 20 epochs, the model achieving the best validation performance was selected, and its final performance was evaluated on a separate, held-out test set consisting of 100,000 events.

\subsection{Performance in the \texorpdfstring{$\tau^+\tau^-$}{tau tau} Channel}

\begin{figure*}[!tbp] 
\centering 
%\vspace{-20pt} 
\includegraphics[width=1\linewidth]{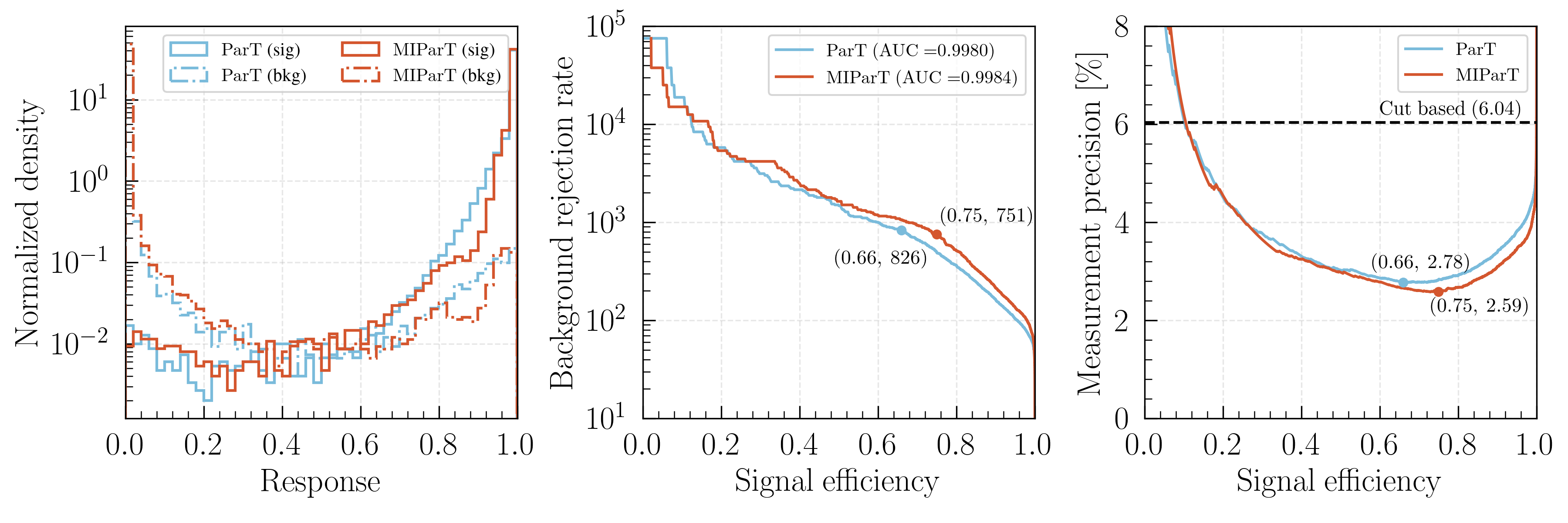}
%\vspace{-35pt} 
\caption{\label{ml_tautau}
The normalized distributions of the classifier response scores for the signal and background events in the testing dataset (left panel), the background rejection rate as a function of signal efficiency (middle panel), and the signal strength measurement precision versus signal efficiency (right panel) in the  $\tau^+\tau^-$ channel.}
\end{figure*}

Fig.~\ref{ml_tautau} presents the normalized distributions of the classifier response scores for signal and background events in the testing dataset (left panel), the background rejection rate as a function of signal efficiency (middle panel), and the signal strength measurement precision versus signal efficiency (right panel) in the $\tau^+\tau^-$ channel.
The signal efficiency corresponds to the fraction of true signal events correctly identified, 
while the background rejection rate indicates the fraction of background events successfully rejected.
\footnote{
The signal efficiency is equivalent to the true positive rate (TPR), 
$\varepsilon_{\mathrm{sig}} = N_{\mathrm{TP}}/(N_{\mathrm{TP}} + N_{\mathrm{FN}})$, 
and the background rejection corresponds to $1/\mathrm{FPR}$, 
where $\mathrm{FPR} = N_{\mathrm{FP}}/(N_{\mathrm{FP}} + N_{\mathrm{TN}})$. 
Here $N_{\mathrm{TP}}$, $N_{\mathrm{FP}}$, $N_{\mathrm{FN}}$, and $N_{\mathrm{TN}}$ denote the number of true positive, false positive, false negative, and true negative events, respectively. A background rejection rate of 200 at signal efficiency is 50\% means that for every 200 true background events, approximately one is misclassified as signal, while half of the true signal events are correctly identified. }
The AUC of ParT and MIParT can reach 0.9980 and 0.9984, respectively. 
Physically, the signal efficiency represents the fraction of true signals correctly identified, while the background rejection rate indicates the fraction of background events successfully rejected. 

For the benchmark point, the $\tau^+\tau^-$ channel signal strength measurement precision reaches 2.78\% using ParT at a signal efficiency of 0.66, which corresponds to a background rejection rate of 826, as shown in the right panels of Fig.~\ref{ml_tautau}. 
A marginal improvement is observed with MIParT, which delivers a precision of 2.59\% at a higher signal efficiency of 0.75 and a higher background rejection rate of 751, indicating its superior performance in this scenario. 
Relative to the 6.04\% measurement precision obtained with the cut-based analysis, the precision is improved by factors of 2.2 and 2.3 when using ParT and MIParT, respectively. 
This improvement is consistent with the findings of Ref.~\cite{Zhu:2025eoe}, 
which demonstrated that a holistic event-level approach combined with Advanced Color Singlet Identification (ACSI) can enhance the expected precision by factors of two to six in processes where the SM-like Higgs boson is produced in association with neutrino or quark pairs and decays into light-flavor quarks or gluons.

A detailed comparison between the cut-based and ML-based analysis for the $\tau^+\tau^-$ channel is presented in Table~\ref{t03}.
The signal efficiency and background rejection rate here are evaluated relative to the events after the basic cut.
The substantial improvement in measurement precision from the ML-based analysis arises from a markedly higher signal efficiency, achieved with only a modest compromise in background rejection rate. 
It is worth noting that, compared to the improvements offered by jet-level ML algorithms such as eXtreme Gradient Boosting, Gradient Boosting Decision Trees, and Deep Neural Networks in our previous work~\cite{Dong:2025orv}, the enhancement from particle-level transformers here is even more pronounced.
This enhancement can be attributed to the efficient use of particle-level features and the superior ability of the ParT and MIParT architectures to extract underlying correlations among them.

\begin{table}[!tbp]
\centering
\caption{Comparison of the cut-based and the ML-based analysis in the $\tau^+\tau^-$ decay channel}
\label{t03}
\begin{tabular*}{\columnwidth}{@{\extracolsep{\fill}}lcccc}
    \toprule
    \multirow{2}{*}{Method} & \multicolumn{1}{c}{Signal} & \multicolumn{1}{c}{Background} & \multirow{2}{*}{Precision} \\
    % \cmidrule(r){2-3}
    & \multicolumn{1}{c}{efficiency} & \multicolumn{1}{c}{rejection rate} & \\
    \midrule
    Cut-based & 0.16 & 2224 & 6.04\% \\ 
    ParT & 0.66 & 826 & 2.78\% \\ 
    MIParT & 0.75 & 751 & 2.59\% \\ 
    \bottomrule
\end{tabular*}
\end{table}

\subsection{Performance in the \texorpdfstring{$b\bar{b}$}{bb} Channel}

\begin{figure*}[!tbp] 
\centering 
%\vspace{-15pt} 
\includegraphics[width=1\linewidth]{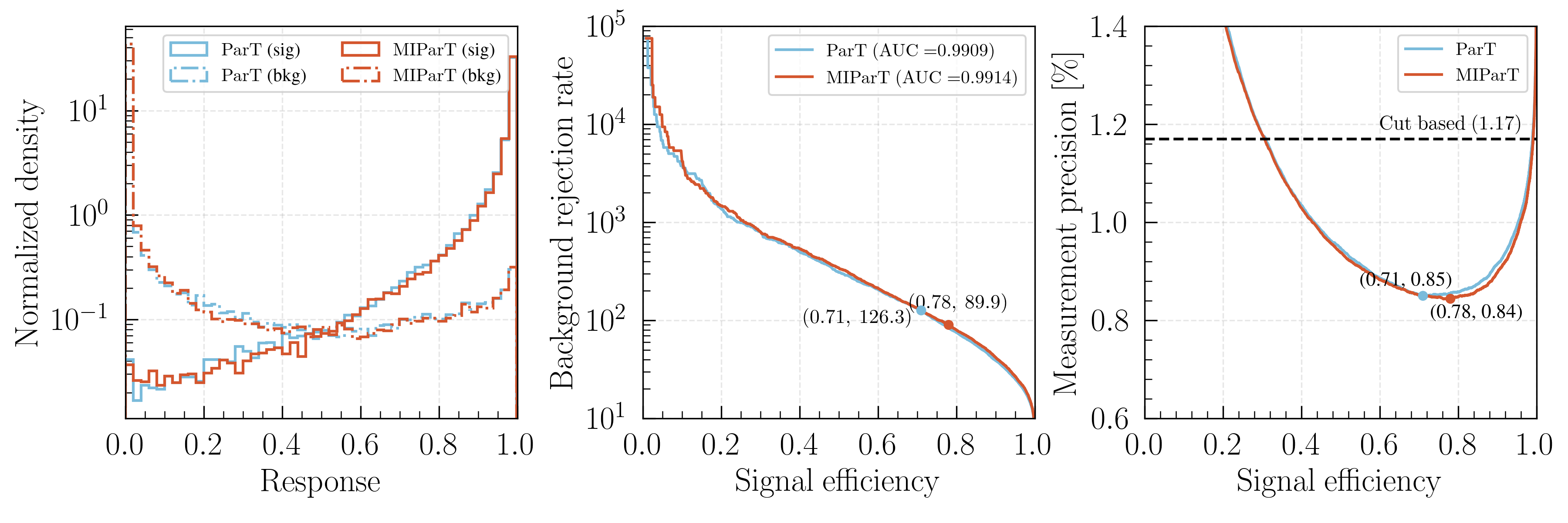}
%\vspace{-35pt} 
\caption{\label{ml_bb}
The normalized distributions of the classifier response scores for the signal and background events in the testing dataset (left panel), the background rejection rate as a function of signal efficiency (middle panel), and the signal strength measurement precision versus signal efficiency (right panel) in the $b\bar{b}$ channel.
}
\end{figure*}

The same ML methodology is applied to the $b\bar{b}$ decay channel. Fig.~\ref{ml_bb} shows the normalized distributions of classifier response scores for signal and background events in the testing dataset (left panel), the background rejection rate as a function of signal efficiency (middle panel), and the signal strength measurement precision versus signal efficiency (right panel) in the $b\bar{b}$ channel.
The AUC values for MIParT and ParT reach 0.9914 and 0.9909, respectively, with MIParT slightly outperforming ParT.

As shown in the middle and right panels of Fig.~\ref{ml_bb}, the ParT model in the $b\bar{b}$ channel achieves a signal strength measurement precision of 0.85\% at a signal efficiency of 0.71, while maintaining a background rejection of 126.3.
In comparison, MIParT achieves a precision of 0.84\% at a slightly higher signal efficiency of 0.78, with a background rejection rate of 89.9. Although MIParT exhibits a marginally higher AUC, the difference in measurement precision between the two models is negligible in this case.
Compared to the cut-based analysis, which yields a precision of 1.17\%, the ML-based analysis improves the measurement precision by a factor of 1.4 in the $b\bar{b}$ channel. A detailed comparison of the cut-based and the ML-based analysis in the $b\bar{b}$ decay channel is provided in Table~\ref{t04}. The ML-based analysis enhances both the signal efficiency and background rejection rate, leading to an overall improvement in measurement precision.

It is worth noting that the improvement achieved by MIParT over ParT is more pronounced in the $\tau^+\tau^-$ channel than in the $b\bar{b}$ channel.
This observation can be understood from the underlying event topology. The $\tau^+\tau^-$ final state involves missing transverse energy and multiple decay modes, resulting in more diverse kinematic and substructure patterns. 
The additional interaction layers in MIParT enable the network to capture these complex inter-particle correlations more effectively. 
In contrast, the $b\bar{b}$ final state carries relatively complete and less ambiguous information; hence, the performance gain from enhanced interaction modeling in MIParT is naturally smaller.

\begin{table}[!tbp]
\centering
\caption{Comparison of the cut-based and the ML-based analysis in the $b\bar{b}$ decay channel}
\label{t04}
\begin{tabular*}{\columnwidth}{@{\extracolsep{\fill}}lcccc}
    \toprule
    \multirow{2}{*}{Method} & \multicolumn{1}{c}{Signal} & \multicolumn{1}{c}{Background} & \multirow{2}{*}{Precision} \\
    % \cmidrule(r){2-3}
    & \multicolumn{1}{c}{efficiency} & \multicolumn{1}{c}{rejection rate} & \\
    \midrule
    Cut-based & 0.69 & 33.7 & 1.17\%  \\ 
    ParT & 0.71 & 126.3 & 0.85\% \\ 
    MIParT & 0.78 & 89.9 & 0.84\% \\ 
    \bottomrule
\end{tabular*}
\end{table}

\subsection{Comparative Discussion}
\begin{figure*}[!tbp]
\centering
\vspace{-10pt} 
\includegraphics[width=0.75\linewidth]{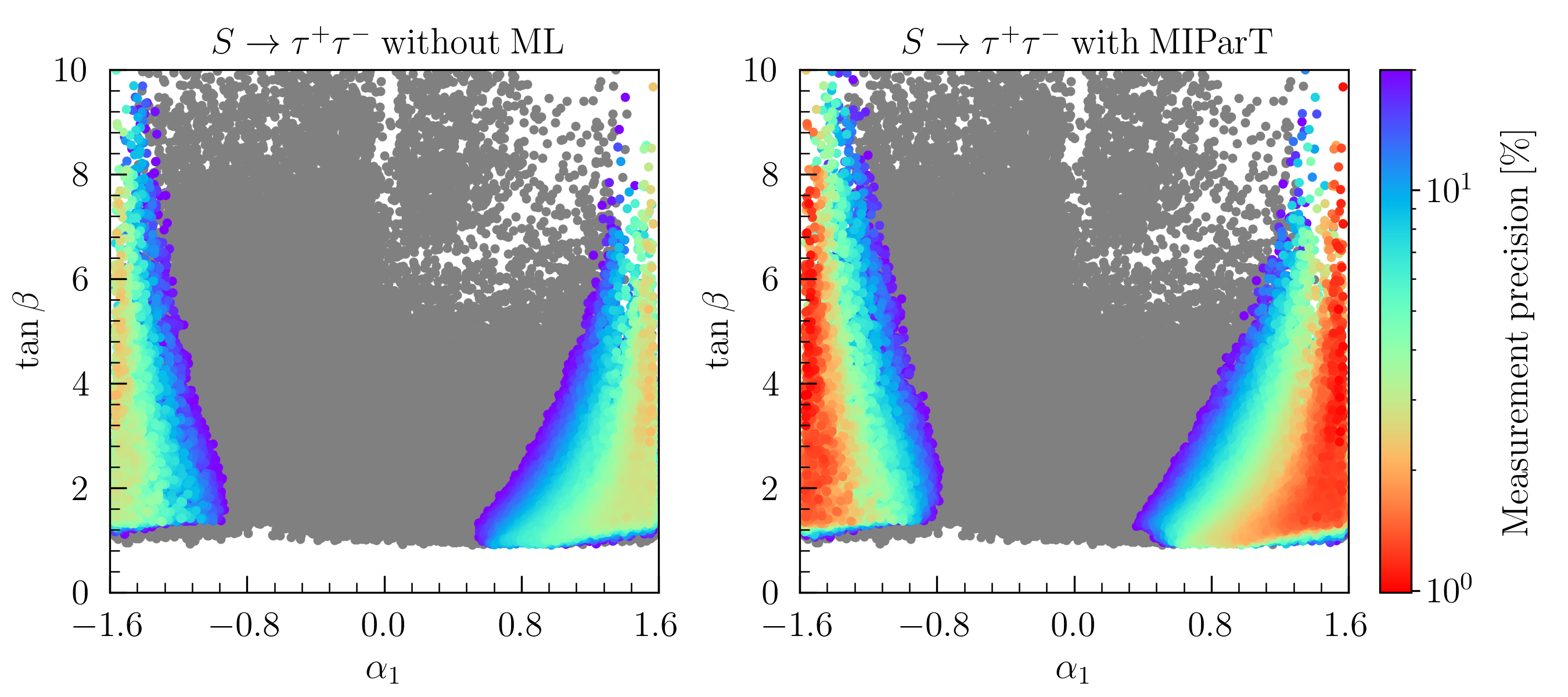}
\vspace{6pt}   
\includegraphics[width=0.75\linewidth]{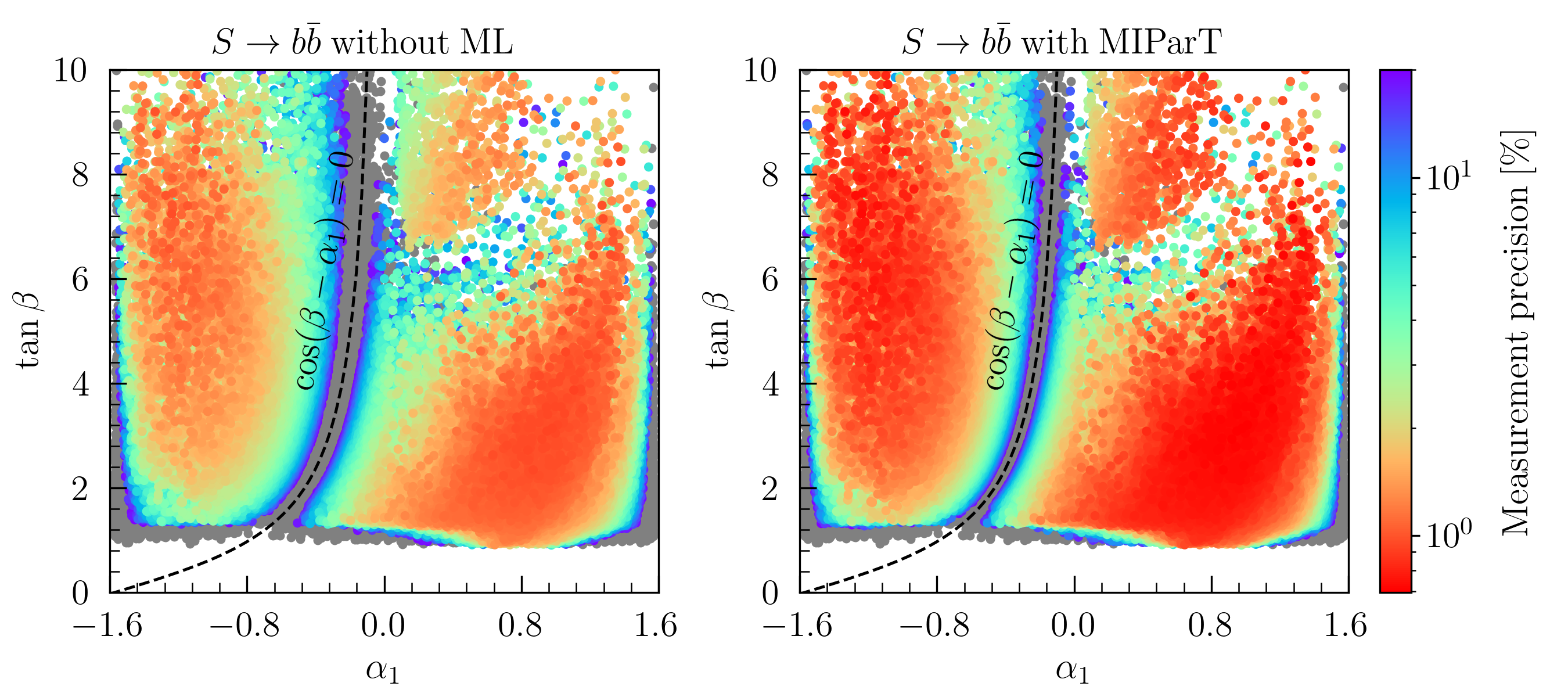}
\vspace{-15pt} 
\caption{The surviving samples in the $\tanb$ versus $\alpha_1$ plane. The color represents the signal strength measurement precision in the $\tau^+\tau^-$ channel without ML (upper left), with MIParT (upper right), in the $b\bar{b}$ channel without ML (lower left), and with MIParT (lower right), based on an integrated luminosity of $L = 20\abm$ at CEPC. Gray points indicate samples that cannot be covered at the 5$\sigma$ signal significance level under the corresponding scenario.}
\label{combined}
\end{figure*}

To assess the coverage of the N2HDM-F parameter space at the 240 GeV CEPC by the cut-based and the ML-based analysis, Fig.~\ref{combined} shows the surviving samples in the $\tan\beta$ versus $\alpha_1$ plane. The color scale indicates the signal strength measurement precision in the $\tau^+\tau^-$ channel without ML (upper left), with MIParT (upper right), in the $b\bar{b}$ channel without ML (lower left), and with MIParT (lower right), assuming an integrated luminosity of $L = 20\abm$ at the CEPC. Gray points represent samples that cannot be covered at the $5\sigma$ signal significance level in the corresponding scenario. The signal significance is computed using the Poisson formula~\cite{Cowan:2010js}
\begin{align}
\mathcal{Z} = \sqrt{2\,\big[(S+B)\ln(1+S/B) - S\big]}\,,
\end{align}
where $S$ and $B$ denote the expected numbers of signal and background events, respectively.

The following conclusions can be drawn from the Fig.~\ref{combined}:
\begin{itemize}
    \item As shown in the upper two panels, for a given \(\alpha_{1}\), samples with larger $\tan\beta$ are more challenging to detect than those with smaller $\tan\beta$.
    Meanwhile, samples with $|\alpha|$ close to $\pi/2$, which corresponds to $|\sin\alpha_1|$ near 1, can be measured with higher precision.
    This observation is consistent with Eq.~\ref{eq_hf}, which indicates that the coupling of $S$ to $\tau$ is proportional to $\sin\alpha_1$ and inversely proportional to $\sin\beta$. The surviving samples with the maximum signal strength in the $\tau^+\tau^-$ channel can be detected with 2.2\% precision without ML, and with 1.0\% precision when using MIParT.
    \item From the lower two panels, one can see that almost all the surviving samples can be covered at the 5$\sigma$ level in the $b\bar{b}$ decay channel except the surviving samples with cos$(\beta-\alpha_1)\approx 0$, as it is the dominant decay channel for most parameter space. In addition, the use of MIParT can further improve the signal strength measurement precision; the maximum can be improved from 0.93\% without ML to 0.69\% with MIParT.
    \item As we expect, the samples with small tan$\beta$ and $|\alpha|$ close to $\pi/2$ are hardly be detected in $b\bar{b}$ channel,  but can be detected in $\tau^+\tau^-$ channel at high measurement precision.
\end{itemize}

\begin{figure*}[!tbp] 
\centering 
%\vspace{-10pt} 
\includegraphics[width=0.7\linewidth]{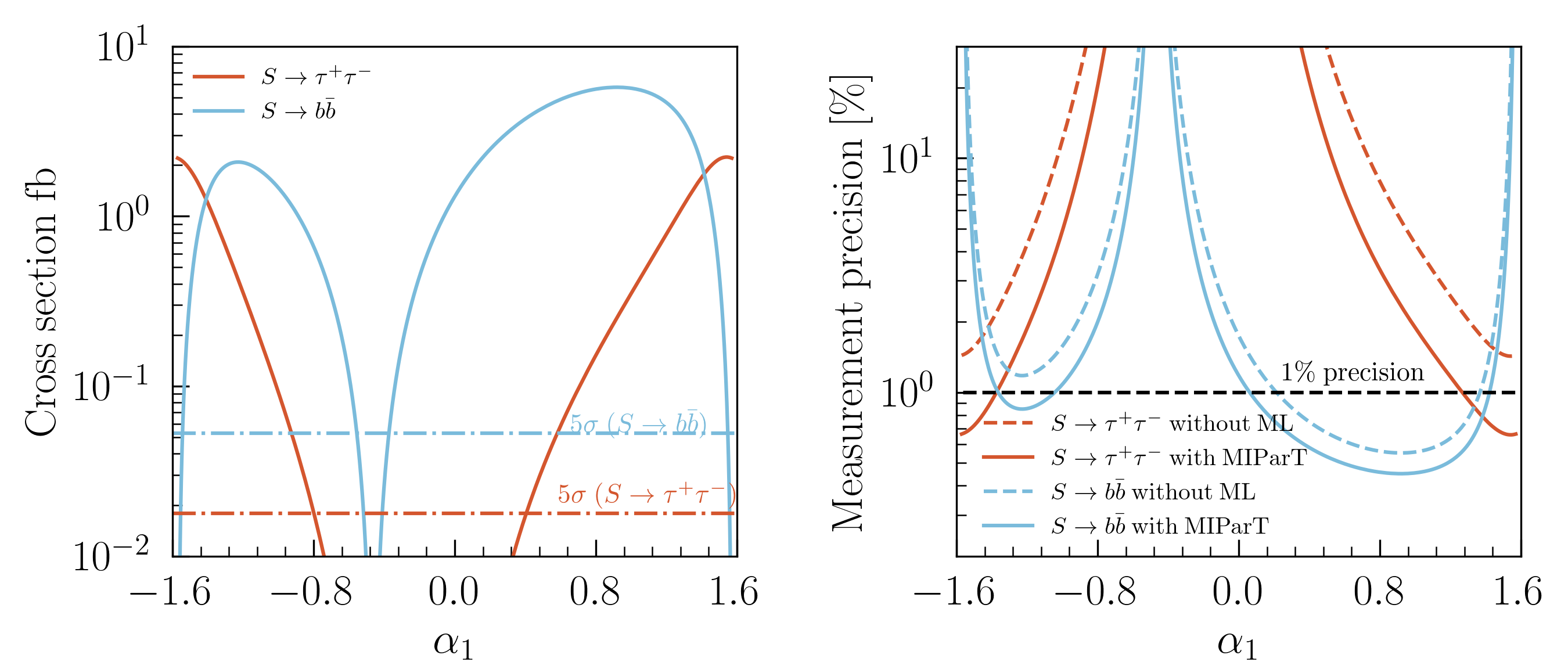}
%\vspace{-15pt} 
\caption{\label{benchmark}
The cross section (left panel) and the signal strength measurement precision (right panel) as a function of the model parameter $\alpha_1$ in the $\tau^+\tau^-$ decay channel (red lines) and the $b\bar{b}$ decay channel (blue lines) for the benchmark points with $\cos\alpha_2 = \sqrt{2}/2$ and $\tan\beta = 2$.
}
\end{figure*}

For a detailed discussion on the improvement in CEPC coverage capacity provided by the ML-based analysis over the cut-based analysis, Fig.~\ref{benchmark} shows the cross section (left panel) and the signal strength measurement precision (right panel) as a function of the model parameter $\alpha_1$ in the $\tau^+\tau^-$ decay channel (red lines) and the $b\bar{b}$ decay channel (blue lines) for the benchmark points with $\cos\alpha_2 = \sqrt{2}/2$ and $\tan\beta = 2$.
Without ML, only samples in the range $0.2 < \alpha_1 < 1.4$ can be measured with 1\% precision in the $b\bar{b}$ decay channel, while no samples satisfy the 1\% precision requirement in the $\tau^+\tau^-$ channel.
After applying MIParT, samples with $\alpha_1 < -1.4$ or $\alpha_1 >1.3$ can be measured at 1\% precision in the $\tau^+\tau^-$ decay channel. Only those in the region $-1 < \alpha_1 < 0$ remain undetectable at this precision level in both the $b\bar{b}$ and $\tau^+\tau^-$ channels.
In terms of $5\sigma$ discovery potential at the 240 GeV CEPC with an integrated luminosity of $L = 20\ \text{ab}^{-1}$, signal regions are considered covered for the cross section exceeds 0.018 fb in the $\tau^+\tau^-$ channel or 0.051 fb in the $b\bar{b}$ channel. The only parameter region inaccessible at this significance is $-0.55 < \alpha_1 < -0.37$.

\begin{figure*}[!tbp] 
\centering 
%\vspace{-10pt} 
\includegraphics[width=0.75\linewidth]{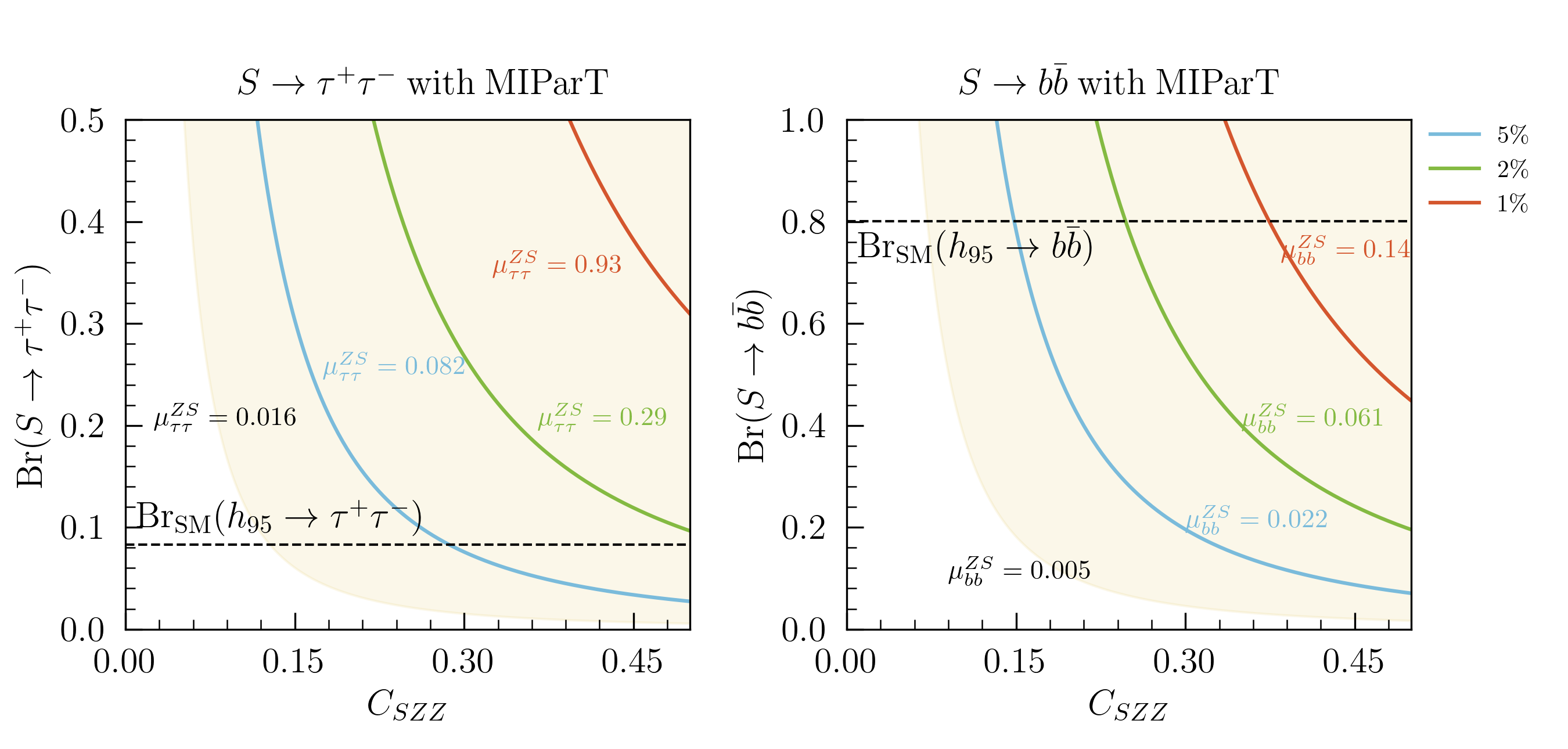}
%\vspace{-15pt} 
\caption{\label{tt_bb_cs}
The CEPC sensitivity in the $\tau^+\tau^-$ and $b\bar{b}$ channels on the Br$(S\to\tau^+\tau^-)$ versus $C_{SZZ}$ plane (left panel) and Br$(S\to b\bar{b})$ versus $C_{SZZ}$ plane (right panel). The shadow area can be covered at the 5$\sigma$ level and the range upper the lines can be detected at the corresponding precision.
}
\end{figure*}

The CEPC signal strength measurement precisions of 5\%, 2\%, and 1\% for the $\tau^+\tau^-$ and $b\bar{b}$ channels are presented in Fig.~\ref{tt_bb_cs} on the $\mathrm{Br}(S \to \tau^+\tau^-)$ versus $C_{SZZ}$ plane (left panel) and the $\mathrm{Br}(S \to b\bar{b})$ versus $C_{SZZ}$ plane (right panel), respectively. The shadow area can be covered at the 5$\sigma$ level and the regions above the contours correspond to parameter spaces accessible at the corresponding precision.
The signal strength in the $\tau^+\tau^-$ or $b\bar{b}$ decay channel from the Higgsstrahlung process is defined as
\begin{align}
\mu_{\tau\tau\ (bb)}^{ZS} = C_{SZZ}^2 \times \frac{\mathrm{Br}(S\to \tau^+\tau^-\ (b\bar{b}))}{\mathrm{Br}_{\mathrm{SM}}(h_{95}\to \tau^+\tau^-\ (b\bar{b}))}.
\end{align}
Based on this definition, the $5\sigma$ discovery thresholds are $\mu_{\tau\tau}^{ZS} > 0.016$ and $\mu_{bb}^{ZS} > 0.005$ for the $\tau^+\tau^-$ and $b\bar{b}$ channels, respectively. Furthermore, the minimum signal strengths required to achieve specific measurement precisions are outlined below:
\begin{itemize}
    \item For a 5\%, 2\%, or 1\% precision in the $\tau^+\tau^-$ channel, $\mu_{\tau\tau}^{ZS}$ should exceed 0.082, 0.29, or 0.93, respectively.
    \item Correspondingly, in the $b\bar{b}$ channel, $\mu_{bb}^{ZS}$ should be greater than 0.022, 0.061, or 0.14.
\end{itemize}

Throughout this work, we report purely statistical uncertainties on the signal strength.
A realistic CEPC analysis will, however, also be subject to experimental and theoretical systematic uncertainties. Primary sources of systematic error include jet reconstruction, jet energy scale and resolution, as well as theoretical uncertainties in the predictions for cross sections and spectral shapes.
Based on studies from the CEPC Technical Design Report~\cite{CEPCStudyGroup:2025kmw}, the total systematic uncertainty is estimated to be on the order of 2\%.
A shift from statistical to systematic uncertainty dominance is expected to occur with our ML method, especially in the $\tau^+ \tau^-$ channel.

In the process of $e^+e^-\to ZS$, as long as the mass of $S$ is fixed, the observables are almost entirely determined by its mass. The coupling constants primarily affect the cross sections and branching ratios, whereas they do not significantly alter the event shapes. Consequently, their impact on the simulation results is relatively limited.
These results then are also applicable to other multi-Higgs models featuring a light Higgs with a mass around 95.5 GeV.

The signal channel with the $Z$ boson decaying to electrons has also been investigated using the same MC simulation.
Due to the comparable tracking and calorimetric performance for electrons and muons in the CEPC baseline detector, the resulting sensitivity is similar.
When the electron and muon channels of the $Z$ boson are combined, the measurement precision can be further improved.
A detailed discussion of the electron channel is omitted here for brevity, but it does not change our main conclusions.
The hadronic $Z\to q\bar q$ decays, which have a much larger branching ratio, are in principle also promising, but require a dedicated analysis with more challenging QCD backgrounds and jet reconstruction.
We therefore leave the exploration of hadronic recoil channels for future work.

\section{\label{sec:conclusion}Conclusions} 

In this work, we have explored the prospects for detecting a light Higgs with a mass of 95\,GeV at future $e^+e^-$ Higgs factories. We take the CEPC as a benchmark, with a center-of-mass energy of $\sqrt{s}=240$ GeV and an integrated luminosity of $L=20~\mathrm{ab}^{-1}$. 
The analysis leverages the ParT architecture and its more-interactive variant, MIParT, which are specifically designed to process particle-level observables and extract their complex correlations.
This advanced capability translates directly into superior measurement precision: on the same MC samples, our ML-based analysis outperforms the cut-based analysis, improving the expected statistical precision by factors of 2.3 and 1.4 in the $\tau^+\tau^-$ and $b\bar{b}$ channels, respectively. 
A shift from statistical to systematic uncertainty dominance is expected to occur with our ML method, especially in the $\tau^+ \tau^-$ channel.
Consequently, the region of the N2HDM-F parameter space accessible to high-precision measurement (1\%) at the CEPC is significantly expanded.
This improvement originates from the efficient exploitation of final-state particle information and the superior capability of transformer architectures to capture global event features and inter-particle dependencies.

Most of the parameter space in the N2HDM-F can be probed at the $5\sigma$ significance level when the $\tau^+\tau^-$ and $b\bar{b}$ channels are combined, except in regions near $\cos(\beta - \alpha_1)\!\approx\!0$, where the coupling of the 95\,GeV scalar $S$ to the $Z$ boson is highly suppressed, leading to small production rates. 
For the surviving points with the largest signal strength, a measurement precision of 2.2\% (0.93\%) can be achieved in the $\tau^+\tau^-$ ($b b$) channel using a cut-based analysis, which improves to 1.0\% (0.69\%) when the MIParT architecture is employed.

To present the results in a model-independent way, we express the sensitivities of the 20$\abm$ CEPC in terms of the signal strength $\mu_{\tau\tau}^{ZS}$ and $\mu_{bb}^{ZS}$.
The ML-based analysis yields the following sensitivity reaches:
\begin{itemize}
    \item \textbf{$\tau^+\tau^-$ channel:} A $5\sigma$ discovery can be achieved for $\mu_{\tau\tau}^{ZS}>1.6\times10^{-2}$, while a 1\% precision measurement is attainable for $\mu_{\tau\tau}^{ZS}>0.93$.
    \item \textbf{$b b$ channel:} A $5\sigma$ discovery can be achieved for $\mu_{bb}^{ZS}>5.0\times10^{-3}$, and a 1\% precision measurement for $\mu_{bb}^{ZS}>0.14$.
\end{itemize}

These results demonstrate that particle-level transformer networks can significantly enhance the sensitivity to light Higgs states, thereby improving the measurement precision at future lepton colliders. 
Although this study is performed using the CEPC baseline detector and luminosity assumptions, the conclusions are broadly applicable to other future $e^+e^-$ Higgs factories such as FCC-ee~\cite{FCC:2018evy, FCC:2018byv} and ILC~\cite{ILC:2013jhg, Behnke:2013xla, Asner:2013psa}. 
Since the ParT/MIParT classifiers rely mainly on kinematic features that depend on the collider energy and on $m_S$, rather than on machine-specific details, we expect the relative improvements over cut-based analysis demonstrated here to hold quite generally for other $e^+e^-$ Higgs factories.
Our work offers a compelling blueprint for applying ML-based analysis to advance the precision Higgs physics program and explore BSM Higgs sectors.

\acknowledgments
This work was supported by the National Natural Science Foundation of China under Grant No. 12275066 and by the startup research funds of Henan University. 
The work of H. Yang was also supported by the National Natural Science Foundation of China under Grant number: W2441004.
The work of M. Ruan was also supported by the National Key Program for S$\&$T Research and Development under Contract No. 2024YFA1610603. 
And the work of K. Wang was also supported by the Open Project of the Shanghai Key Laboratory of Particle Physics and Cosmology under Grant No. 22DZ2229013-3.

\appendix

\section{Details of the ParT and MIParT implementations}
\label{app:ml}

We briefly summarise the main hyperparameters and setup for our ParT and MIParT implementations. Both architectures follow the original designs presented in Refs.~\cite{Qu:2022mxj} (ParT) and \cite{Wu:2024thh} (MIParT), with the following specific choices:

\begin{itemize}
    \item \textbf{Embedding dimension} \(d_{\text{model}}\): 128 for ParT, 64 for MIParT.
    \item \textbf{Transformer blocks}: ParT uses 8 Particle- and 2 Class-attention layers; MIParT uses 5 MI-Particle, 5 Particle, and 2 Class-attention layers.
    \item \textbf{Attention heads}: \(N_{\text{heads}} = 8\) for all attention layers.
    \item \textbf{Feed-forward hidden dimension} \(d_{\text{ff}}\): 512 for ParT, 256 for MIParT.
    \item \textbf{Dropout rate}: 0.1.
    \item \textbf{Optimiser}: RAdam with an initial learning rate of \(1\times10^{-3}\).
    \item \textbf{Loss function}: Cross entropy with equal class weighting to mitigate class imbalance.
    \item \textbf{Batch size}: 126. 
    \item \textbf{Training epochs}: 20.
\end{itemize}

The dataset was randomly split into 90\% for training and 10\% for validation. An independent sample of 100k events was held out for testing. No signs of overfitting were observed; the performance on the test set was consistent with that on the validation set.

Further architectural details, along with the complete training code, can be found in the original references and the associated GitHub home page :
\href{https://github.com/hqucms/weaver-core}{weaver}, 
\href{https://github.com/jet-universe/particle_transformer}{ParT}, and
\href{https://github.com/USST-HEP/MIParT}{MIParT}.

%\newpage
% whether or not they are referenced in the text. This is an appropriate
% \nocite{*}

% \bibliographystyle{apsrev4-2}
\bibliographystyle{apsrev4-1}
\bibliography{apssamp}% Produces the bibliography via BibTeX.

@article{CMS:2012qbp,
    author = "Chatrchyan, Serguei and others",
    collaboration = "CMS",
    title = "{Observation of a New Boson at a Mass of 125 GeV with the CMS Experiment at the LHC}",
    eprint = "1207.7235",
    archivePrefix = "arXiv",
    primaryClass = "hep-ex",
    reportNumber = "CMS-HIG-12-028, CERN-PH-EP-2012-220",
    doi = "10.1016/j.physletb.2012.08.021",
    journal = "Phys. Lett. B",
    volume = "716",
    pages = "30--61",
    year = "2012"
}

@article{ATLAS:2012yve,
    author = "Aad, Georges and others",
    collaboration = "ATLAS",
    title = "{Observation of a new particle in the search for the Standard Model Higgs boson with the ATLAS detector at the LHC}",
    eprint = "1207.7214",
    archivePrefix = "arXiv",
    primaryClass = "hep-ex",
    reportNumber = "CERN-PH-EP-2012-218",
    doi = "10.1016/j.physletb.2012.08.020",
    journal = "Phys. Lett. B",
    volume = "716",
    pages = "1--29",
    year = "2012"
}

@article{Kajantie:1995kf,
    author = "Kajantie, K. and Laine, M. and Rummukainen, K. and Shaposhnikov, Mikhail E.",
    title = "{The Electroweak phase transition: A Nonperturbative analysis}",
    eprint = "hep-lat/9510020",
    archivePrefix = "arXiv",
    reportNumber = "CERN-TH-95-263, HD-THEP-95-44, HU-TFT-95-57, IUHET-318",
    doi = "10.1016/0550-3213(96)00052-1",
    journal = "Nucl. Phys. B",
    volume = "466",
    pages = "189--258",
    year = "1996"
}

@article{Farrar:1993sp,
    author = "Farrar, Glennys R. and Shaposhnikov, M. E.",
    title = "{Baryon asymmetry of the universe in the minimal Standard Model}",
    eprint = "hep-ph/9305274",
    archivePrefix = "arXiv",
    reportNumber = "RU-93-10, CERN-TH-6729-92",
    doi = "10.1103/PhysRevLett.70.2833",
    journal = "Phys. Rev. Lett.",
    volume = "70",
    pages = "2833--2836",
    year = "1993",
    note = "[Erratum: Phys.Rev.Lett. 71, 210 (1993)]"
}

@article{Gavela:1993ts,
    author = "Gavela, M. B. and Hernandez, P. and Orloff, J. and Pene, O.",
    title = "{Standard model CP violation and baryon asymmetry}",
    eprint = "hep-ph/9312215",
    archivePrefix = "arXiv",
    reportNumber = "CERN-TH-7081-93, LPTHE-ORSAY-93-48, HUTP-93-A036, HD-THEP-93-45",
    doi = "10.1142/S0217732394000629",
    journal = "Mod. Phys. Lett. A",
    volume = "9",
    pages = "795--810",
    year = "1994"
}

@article{Sakharov:1967dj,
    author = "Sakharov, A. D.",
    title = "{Violation of CP Invariance, C asymmetry, and baryon asymmetry of the universe}",
    doi = "10.1070/PU1991v034n05ABEH002497",
    journal = "Pisma Zh. Eksp. Teor. Fiz.",
    volume = "5",
    pages = "32--35",
    year = "1967"
}

@article{Huet:1994jb,
    author = "Huet, Patrick and Sather, Eric",
    title = "{Electroweak baryogenesis and standard model CP violation}",
    eprint = "hep-ph/9404302",
    archivePrefix = "arXiv",
    reportNumber = "SLAC-PUB-6479",
    doi = "10.1103/PhysRevD.51.379",
    journal = "Phys. Rev. D",
    volume = "51",
    pages = "379--394",
    year = "1995"
}

@inproceedings{Rummukainen:1998nu,
    author = "Rummukainen, K. and Kajantie, K. and Laine, M. and Shaposhnikov, Mikhail E. and Tsypin, M.",
    title = "{The Universal properties of the electroweak phase transition}",
    booktitle = "{5th International Workshop on Thermal Field Theories and Their Applications}",
    eprint = "hep-ph/9809435",
    archivePrefix = "arXiv",
    reportNumber = "NORDITA-98-69-HE",
    month = "9",
    year = "1998"
}

@article{Anisha:2022hgv,
    author = {Anisha and Biermann, Lisa and Englert, Christoph and M{\"u}hlleitner, Margarete},
    title = "{Two Higgs doublets, effective interactions and a strong first-order electroweak phase transition}",
    eprint = "2204.06966",
    archivePrefix = "arXiv",
    primaryClass = "hep-ph",
    doi = "10.1007/JHEP08(2022)091",
    journal = "JHEP",
    volume = "08",
    pages = "091",
    year = "2022"
}

@article{LEPWorkingGroupforHiggsbosonsearches:2003ing,
    author = "Barate, R. and others",
    collaboration = "LEP Working Group for Higgs boson searches, ALEPH, DELPHI, L3, OPAL",
    title = "{Search for the standard model Higgs boson at LEP}",
    eprint = "hep-ex/0306033",
    archivePrefix = "arXiv",
    reportNumber = "CERN-EP-2003-011",
    doi = "10.1016/S0370-2693(03)00614-2",
    journal = "Phys. Lett. B",
    volume = "565",
    pages = "61--75",
    year = "2003"
}

@article{CMS:2018cyk,
    author = "Sirunyan, Albert M and others",
    collaboration = "CMS",
    title = "{Search for a standard model-like Higgs boson in the mass range between 70 and 110 GeV in the diphoton final state in proton-proton collisions at $\sqrt{s}=$ 8 and 13 TeV}",
    eprint = "1811.08459",
    archivePrefix = "arXiv",
    primaryClass = "hep-ex",
    reportNumber = "CMS-HIG-17-013, CERN-EP-2018-207",
    doi = "10.1016/j.physletb.2019.03.064",
    journal = "Phys. Lett. B",
    volume = "793",
    pages = "320--347",
    year = "2019"
}

@article{CMS:2022goy,
    author = "Tumasyan, Armen and others",
    collaboration = "CMS",
    title = "{Searches for additional Higgs bosons and for vector leptoquarks in $\tau\tau$ final states in proton-proton collisions at $\sqrt{s}$ = 13 TeV}",
    eprint = "2208.02717",
    archivePrefix = "arXiv",
    primaryClass = "hep-ex",
    reportNumber = "CMS-HIG-21-001, CERN-EP-2022-137",
    doi = "10.1007/JHEP07(2023)073",
    journal = "JHEP",
    volume = "07",
    pages = "073",
    year = "2023"
}

@article{CMS:2024yhz,
    author = "Hayrapetyan, Aram and others",
    collaboration = "CMS",
    title = "{Search for a standard model-like Higgs boson in the mass range between 70 and 110 GeV in the diphoton final state in proton-proton collisions at s=13TeV}",
    eprint = "2405.18149",
    archivePrefix = "arXiv",
    primaryClass = "hep-ex",
    reportNumber = "CMS-HIG-20-002, CERN-EP-2024-088",
    doi = "10.1016/j.physletb.2024.139067",
    journal = "Phys. Lett. B",
    volume = "860",
    pages = "139067",
    year = "2025"
}

@article{ATLAS:2024bjr,
    author = "Aad, Georges and others",
    collaboration = "ATLAS",
    title = "{Search for diphoton resonances in the 66 to 110 GeV mass range using pp collisions at $ \sqrt{s} $ = 13 TeV with the ATLAS detector}",
    eprint = "2407.07546",
    archivePrefix = "arXiv",
    primaryClass = "hep-ex",
    reportNumber = "CERN-EP-2024-166",
    doi = "10.1007/JHEP01(2025)053",
    journal = "JHEP",
    volume = "01",
    pages = "053",
    year = "2025"
}

@article{Sharma:2024vhv,
    author = "Sharma, Pramod and Mulaudzi, Anza-Tshilidzi and Mosala, Karabo and Mathaha, Thuso and Kumar, Mukesh and Mellado, Bruce and Crivellin, Andreas and Titov, Maxim and Ruan, Manqi and Fang, Yaquan",
    title = "{Discovery potential of future electron-positron colliders for a 95 GeV scalar}",
    eprint = "2407.16806",
    archivePrefix = "arXiv",
    primaryClass = "hep-ph",
    doi = "10.1016/j.physletb.2025.139953",
    journal = "Phys. Lett. B",
    volume = "870",
    pages = "139953",
    year = "2025"
}

@article{Wang:2024bkg,
    author = "Wang, Kun and Zhu, Jingya",
    title = "{95 GeV light Higgs in the top-pair-associated diphoton channel at the LHC in the minimal dilaton model*}",
    eprint = "2402.11232",
    archivePrefix = "arXiv",
    primaryClass = "hep-ph",
    doi = "10.1088/1674-1137/ad4268",
    journal = "Chin. Phys. C",
    volume = "48",
    number = "7",
    pages = "073105",
    year = "2024"
}

@article{Dutta:2023cig,
    author = "Dutta, Juhi and Lahiri, Jayita and Li, Cheng and Moortgat-Pick, Gudrid and Tabira, Sheikh Farah and Ziegler, Julia Anabell",
    title = "{Dark matter phenomenology in 2HDMS in light of the 95 GeV excess}",
    eprint = "2308.05653",
    archivePrefix = "arXiv",
    primaryClass = "hep-ph",
    reportNumber = "DESY-23-114",
    doi = "10.1140/epjc/s10052-024-13176-9",
    journal = "Eur. Phys. J. C",
    volume = "84",
    number = "9",
    pages = "926",
    year = "2024"
}

@article{Dong:2024ipo,
    author = "Dong, Yabo and Wang, Kun and Zhu, Jingya",
    title = "{Probing a type I 2HDM light Higgs boson in the top-pair-associated diphoton channel}",
    eprint = "2410.13636",
    archivePrefix = "arXiv",
    primaryClass = "hep-ph",
    doi = "10.1103/2ss7-ygnb",
    journal = "Phys. Rev. D",
    volume = "112",
    number = "5",
    pages = "055013",
    year = "2025"
}

@article{Xu:2025vmy,
    author = "Xu, Haotian and Wang, Yufei and Han, Xiao-Fang and Wang, Lei",
    title = "{95 GeV Higgs boson and nano-Hertz gravitational waves from domain walls in the next-to-two-Higgs-doublet model*}",
    eprint = "2505.03592",
    archivePrefix = "arXiv",
    primaryClass = "hep-ph",
    doi = "10.1088/1674-1137/ae2082",
    journal = "Chin. Phys.",
    volume = "50",
    number = "1",
    pages = "013108",
    year = "2026"
}

@article{Dong:2025orv,
    author = "Dong, Yabo and Ruan, Manqi and Wang, Kun and Yang, Haijun and Zhu, Jingya",
    title = "{Testing a 95 GeV Scalar at the CEPC with Machine Learning}",
    eprint = "2506.21454",
    archivePrefix = "arXiv",
    primaryClass = "hep-ph",
    month = "6",
    year = "2025"
}

@article{CMS:2007sch,
    author = "Bayatian, G. L. and others",
    collaboration = "CMS",
    title = "{CMS technical design report, volume II: Physics performance}",
    reportNumber = "CERN-LHCC-2006-021, CMS-TDR-008-2",
    doi = "10.1088/0954-3899/34/6/S01",
    journal = "J. Phys. G",
    volume = "34",
    number = "6",
    pages = "995--1579",
    year = "2007"
}

@article{ATLAS:2010ojh,
    author = "Capeans, M. and Darbo, G. and Einsweiler, K. and Elsing, M. and Flick, T. and Garcia-Sciveres, M. and Gemme, C. and Pernegger, H. and Rohne, O. and Vuillermet, R.",
    collaboration = "ATLAS",
    title = "{ATLAS Insertable B-Layer Technical Design Report}",
    reportNumber = "CERN-LHCC-2010-013, ATLAS-TDR-19",
    month = "9",
    year = "2010"
}

@article{FCC:2018evy,
    author = "Abada, A. and others",
    collaboration = "FCC",
    title = "{FCC-ee: The Lepton Collider}: {Future Circular Collider Conceptual Design Report Volume 2}",
    reportNumber = "CERN-ACC-2018-0057",
    doi = "10.1140/epjst/e2019-900045-4",
    journal = "Eur. Phys. J. ST",
    volume = "228",
    number = "2",
    pages = "261--623",
    year = "2019"
}

@article{FCC:2018byv,
    author = "Abada, A. and others",
    collaboration = "FCC",
    title = "{FCC Physics Opportunities}: {Future Circular Collider Conceptual Design Report Volume 1}",
    reportNumber = "CERN-ACC-2018-0056",
    doi = "10.1140/epjc/s10052-019-6904-3",
    journal = "Eur. Phys. J. C",
    volume = "79",
    number = "6",
    pages = "474",
    year = "2019"
}

@article{ILC:2013jhg,
    editor = "Baer, Howard and others",
    collaboration = "ILC",
    title = "{The International Linear Collider Technical Design Report - Volume 2: Physics}",
    eprint = "1306.6352",
    archivePrefix = "arXiv",
    primaryClass = "hep-ph",
    reportNumber = "ILC-REPORT-2013-040, ANL-HEP-TR-13-20, BNL-100603-2013-IR, IRFU-13-59, CERN-ATS-2013-037, COCKCROFT-13-10, CLNS-13-2085, DESY-13-062, FERMILAB-TM-2554, IHEP-AC-ILC-2013-001, INFN-13-04-LNF, JAI-2013-001, JINR-E9-2013-35, JLAB-R-2013-01, KEK-REPORT-2013-1, KNU-CHEP-ILC-2013-1, LLNL-TR-635539, SLAC-R-1004, ILC-HIGRADE-REPORT-2013-003",
    month = "6",
    year = "2013"
}

@article{Behnke:2013xla,
    editor = "Behnke, Ties and Brau, James E. and Foster, Brian and Fuster, Juan and Harrison, Mike and Paterson, James McEwan and Peskin, Michael and Stanitzki, Marcel and Walker, Nicholas and Yamamoto, Hitoshi",
    title = "{The International Linear Collider Technical Design Report - Volume 1: Executive Summary}",
    eprint = "1306.6327",
    archivePrefix = "arXiv",
    primaryClass = "physics.acc-ph",
    reportNumber = "ILC-REPORT-2013-040, ANL-HEP-TR-13-20, BNL-100603-2013-IR, IRFU-13-59, CERN-ATS-2013-037, COCKCROFT-13-10, CLNS-13-2085, DESY-13-062, FERMILAB-TM-2554, IHEP-AC-ILC-2013-001, INFN-13-04-LNF, JAI-2013-001, JINR-E9-2013-35, JLAB-R-2013-01, KEK-REPORT-2013-1, KNU-CHEP-ILC-2013-1, LLNL-TR-635539, SLAC-R-1004, ILC-HIGRADE-REPORT-2013-003",
    month = "6",
    year = "2013"
}

@inproceedings{Asner:2013psa,
    author = "Asner, D. M. and others",
    title = "{ILC Higgs White Paper}",
    booktitle = "{Snowmass 2013}: {Snowmass on the Mississippi}",
    eprint = "1310.0763",
    archivePrefix = "arXiv",
    primaryClass = "hep-ph",
    month = "10",
    year = "2013"
}

@article{CEPCStudyGroup:2018ghi,
    author = "Dong, Mingyi and others",
    editor = "Guimar{\~a}es da Costa, Jo{\~a}o Barreiro and others",
    collaboration = "CEPC Study Group",
    title = "{CEPC Conceptual Design Report: Volume 2 - Physics {\&} Detector}",
    eprint = "1811.10545",
    archivePrefix = "arXiv",
    primaryClass = "hep-ex",
    reportNumber = "IHEP-CEPC-DR-2018-02, IHEP-EP-2018-01, IHEP-TH-2018-01",
    month = "11",
    year = "2018"
}

@article{CEPCStudyGroup:2018rmc,
    collaboration = "CEPC Study Group",
    title = "{CEPC Conceptual Design Report: Volume 1 - Accelerator}",
    eprint = "1809.00285",
    archivePrefix = "arXiv",
    primaryClass = "physics.acc-ph",
    reportNumber = "IHEP-CEPC-DR-2018-01, IHEP-AC-2018-01",
    month = "9",
    year = "2018"
}

@article{An:2018dwb,
    author = "An, Fenfen and others",
    title = "{Precision Higgs physics at the CEPC}",
    eprint = "1810.09037",
    archivePrefix = "arXiv",
    primaryClass = "hep-ex",
    reportNumber = "FERMILAB-PUB-18-573-T",
    doi = "10.1088/1674-1137/43/4/043002",
    journal = "Chin. Phys. C",
    volume = "43",
    number = "4",
    pages = "043002",
    year = "2019"
}

@article{CEPCStudyGroup:2023quu,
    author = "Abdallah, Waleed and others",
    collaboration = "CEPC Study Group",
    title = "{CEPC Technical Design Report: Accelerator}",
    eprint = "2312.14363",
    archivePrefix = "arXiv",
    primaryClass = "physics.acc-ph",
    reportNumber = "IHEP-CEPC-DR-2023-01, IHEP-AC-2023-01",
    doi = "10.1007/s41605-024-00463-y",
    journal = "Radiat. Detect. Technol. Methods",
    volume = "8",
    number = "1",
    pages = "1--1105",
    year = "2024",
    note = "[Erratum: Radiat.Detect.Technol.Methods 9, 184--192 (2025)]"
}

@article{Komiske:2018cqr,
    author = "Komiske, Patrick T. and Metodiev, Eric M. and Thaler, Jesse",
    title = "{Energy Flow Networks: Deep Sets for Particle Jets}",
    eprint = "1810.05165",
    archivePrefix = "arXiv",
    primaryClass = "hep-ph",
    reportNumber = "MIT-CTP 5064",
    doi = "10.1007/JHEP01(2019)121",
    journal = "JHEP",
    volume = "01",
    pages = "121",
    year = "2019"
}

@article{Qu:2019gqs,
    author = "Qu, Huilin and Gouskos, Loukas",
    title = "{ParticleNet: Jet Tagging via Particle Clouds}",
    eprint = "1902.08570",
    archivePrefix = "arXiv",
    primaryClass = "hep-ph",
    doi = "10.1103/PhysRevD.101.056019",
    journal = "Phys. Rev. D",
    volume = "101",
    number = "5",
    pages = "056019",
    year = "2020"
}

@article{Mikuni:2020wpr,
    author = "Mikuni, Vinicius and Canelli, Florencia",
    title = "{ABCNet: An attention-based method for particle tagging}",
    eprint = "2001.05311",
    archivePrefix = "arXiv",
    primaryClass = "physics.data-an",
    reportNumber = "135",
    doi = "10.1140/epjp/s13360-020-00497-3",
    journal = "Eur. Phys. J. Plus",
    volume = "135",
    number = "6",
    pages = "463",
    year = "2020"
}

@article{Mikuni:2021pou,
    author = "Mikuni, Vinicius and Canelli, Florencia",
    title = "{Point cloud transformers applied to collider physics}",
    eprint = "2102.05073",
    archivePrefix = "arXiv",
    primaryClass = "physics.data-an",
    doi = "10.1088/2632-2153/ac07f6",
    journal = "Mach. Learn. Sci. Tech.",
    volume = "2",
    number = "3",
    pages = "035027",
    year = "2021"
}

@article{Gong:2022lye,
    author = "Gong, Shiqi and Meng, Qi and Zhang, Jue and Qu, Huilin and Li, Congqiao and Qian, Sitian and Du, Weitao and Ma, Zhi-Ming and Liu, Tie-Yan",
    title = "{An efficient Lorentz equivariant graph neural network for jet tagging}",
    eprint = "2201.08187",
    archivePrefix = "arXiv",
    primaryClass = "hep-ph",
    doi = "10.1007/JHEP07(2022)030",
    journal = "JHEP",
    volume = "07",
    pages = "030",
    year = "2022"
}

@article{Qu:2022mxj,
    author = "Qu, Huilin and Li, Congqiao and Qian, Sitian",
    title = "{Particle Transformer for Jet Tagging}",
    eprint = "2202.03772",
    archivePrefix = "arXiv",
    primaryClass = "hep-ph",
    month = "2",
    year = "2022"
}

@article{Wu:2024thh,
    author = "Wu, Yifan and Wang, Kun and Li, Congqiao and Qu, Huilin and Zhu, Jingya",
    title = "{Jet tagging with more-interaction particle transformer*}",
    eprint = "2407.08682",
    archivePrefix = "arXiv",
    primaryClass = "hep-ph",
    doi = "10.1088/1674-1137/ad7f3d",
    journal = "Chin. Phys. C",
    volume = "49",
    number = "1",
    pages = "013110",
    year = "2025"
}

@article{Kundu:2019nqo,
    author = "Kundu, Anirban and Maharana, Suvam and Mondal, Poulami",
    title = "{A 96 GeV scalar tagged to dark matter models}",
    eprint = "1907.12808",
    archivePrefix = "arXiv",
    primaryClass = "hep-ph",
    doi = "10.1016/j.nuclphysb.2020.115057",
    journal = "Nucl. Phys. B",
    volume = "955",
    pages = "115057",
    year = "2020"
}

@article{Liu:2018ryo,
    author = "Liu, Lijia and Qiao, Haoxue and Wang, Kun and Zhu, Jingya",
    title = "{A Light Scalar in the Minimal Dilaton Model in Light of LHC Constraints}",
    eprint = "1812.00107",
    archivePrefix = "arXiv",
    primaryClass = "hep-ph",
    doi = "10.1088/1674-1137/43/2/023104",
    journal = "Chin. Phys. C",
    volume = "43",
    number = "2",
    pages = "023104",
    year = "2019"
}

@article{Borah:2023hqw,
    author = "Borah, Debasish and Mahapatra, Satyabrata and Paul, Partha Kumar and Sahu, Narendra",
    title = "{Scotogenic U(1)L{\ensuremath{\mu}}-L{\ensuremath{\tau}} origin of (g-2){\ensuremath{\mu}}, W-mass anomaly and 95~GeV excess}",
    eprint = "2310.11953",
    archivePrefix = "arXiv",
    primaryClass = "hep-ph",
    doi = "10.1103/PhysRevD.109.055021",
    journal = "Phys. Rev. D",
    volume = "109",
    number = "5",
    pages = "055021",
    year = "2024"
}

@article{Abdelalim:2020xfk,
    author = "Abdelalim, Ahmed Ali and Das, Biswaranjan and Khalil, Shaaban and Moretti, Stefano",
    title = "{Di-photon decay of a light Higgs state in the BLSSM}",
    eprint = "2012.04952",
    archivePrefix = "arXiv",
    primaryClass = "hep-ph",
    doi = "10.1016/j.nuclphysb.2022.116013",
    journal = "Nucl. Phys. B",
    volume = "985",
    pages = "116013",
    year = "2022"
}

@article{Ashanujjaman:2023etj,
    author = "Ashanujjaman, Saiyad and Banik, Sumit and Coloretti, Guglielmo and Crivellin, Andreas and Mellado, Bruce and Mulaudzi, Anza-Tshilidzi",
    title = "{SU(2)L triplet scalar as the origin of the 95~GeV excess?}",
    eprint = "2306.15722",
    archivePrefix = "arXiv",
    primaryClass = "hep-ph",
    reportNumber = "PSI-PR-23-20, ZU-TH 28/23, ICPP-71",
    doi = "10.1103/PhysRevD.108.L091704",
    journal = "Phys. Rev. D",
    volume = "108",
    number = "9",
    pages = "L091704",
    year = "2023"
}

@article{Banik:2023ecr,
    author = "Banik, Sumit and Crivellin, Andreas and Iguro, Syuhei and Kitahara, Teppei",
    title = "{Asymmetric di-Higgs signals of the next-to-minimal 2HDM with a U(1) symmetry}",
    eprint = "2303.11351",
    archivePrefix = "arXiv",
    primaryClass = "hep-ph",
    reportNumber = "PSI-PR-23-7, ZU-TH 15/23, P3H-23-015, TTP23-010, KEK-TH-2506",
    doi = "10.1103/PhysRevD.108.075011",
    journal = "Phys. Rev. D",
    volume = "108",
    number = "7",
    pages = "075011",
    year = "2023"
}

@article{Coloretti:2023yyq,
    author = "Coloretti, Guglielmo and Crivellin, Andreas and Mellado, Bruce",
    title = "{Combined explanation of LHC multilepton, diphoton, and top-quark excesses}",
    eprint = "2312.17314",
    archivePrefix = "arXiv",
    primaryClass = "hep-ph",
    reportNumber = "PSI-PR-24-01, ZU-TH 01/24, ICPP-78",
    doi = "10.1103/PhysRevD.110.073001",
    journal = "Phys. Rev. D",
    volume = "110",
    number = "7",
    pages = "073001",
    year = "2024"
}

@article{Maniatis:2023aww,
    author = "Maniatis, M. and Nachtmann, O.",
    title = "{CMS results for the {\ensuremath{\gamma}}{\ensuremath{\gamma}} production at the LHC: Do they give a hint for a Higgs boson of the maximally CP symmetric two-Higgs-doublet model?}",
    eprint = "2309.04869",
    archivePrefix = "arXiv",
    primaryClass = "hep-ph",
    doi = "10.1103/PhysRevD.111.055009",
    journal = "Phys. Rev. D",
    volume = "111",
    number = "5",
    pages = "055009",
    year = "2025"
}

@article{Wang:2018vxp,
    author = "Wang, Kun and Wang, Fei and Zhu, Jingya and Jie, Quanlin",
    title = "{The semi-constrained NMSSM in light of muon g-2, LHC, and dark matter constraints}",
    eprint = "1811.04435",
    archivePrefix = "arXiv",
    primaryClass = "hep-ph",
    doi = "10.1088/1674-1137/42/10/103109",
    journal = "Chin. Phys. C",
    volume = "42",
    number = "10",
    pages = "103109--103109",
    year = "2018"
}

@article{Cao:2019ofo,
    author = "Cao, Junjie and Jia, Xinglong and Yue, Yuanfang and Zhou, Haijing and Zhu, Pengxuan",
    title = "{96 GeV diphoton excess in seesaw extensions of the natural NMSSM}",
    eprint = "1908.07206",
    archivePrefix = "arXiv",
    primaryClass = "hep-ph",
    doi = "10.1103/PhysRevD.101.055008",
    journal = "Phys. Rev. D",
    volume = "101",
    number = "5",
    pages = "055008",
    year = "2020"
}

@article{Li:2023kbf,
    author = "Li, Weichao and Qiao, Haoxue and Wang, Kun and Zhu, Jingya",
    title = "{Light dark matter confronted with the 95 GeV diphoton excess}",
    eprint = "2312.17599",
    archivePrefix = "arXiv",
    primaryClass = "hep-ph",
    month = "12",
    year = "2023"
}

@article{Wang:2022okq,
    author = "Wang, Chu and Tao, Jun-Quan and Shahzad, M. Aamir and Chen, Guo-Ming and Gascon-Shotkin, S.",
    title = "{Search for a lighter neutral custodial fiveplet scalar in the Georgi-Machacek model *}",
    eprint = "2204.09198",
    archivePrefix = "arXiv",
    primaryClass = "hep-ph",
    reportNumber = "46(8): 083107",
    doi = "10.1088/1674-1137/ac6cd3",
    journal = "Chin. Phys. C",
    volume = "46",
    number = "8",
    pages = "083107",
    year = "2022"
}

@article{Aguilar-Saavedra:2020wrj,
    author = "Aguilar-Saavedra, Juan Antonio and Joaquim, Filipe Rafael",
    title = "{Multiphoton signals of a (96 GeV?) stealth boson}",
    eprint = "2002.07697",
    archivePrefix = "arXiv",
    primaryClass = "hep-ph",
    reportNumber = "IFT-UAM/CSIC-19-153, CFTP/20-002",
    doi = "10.1140/epjc/s10052-020-7952-4",
    journal = "Eur. Phys. J. C",
    volume = "80",
    number = "5",
    pages = "403",
    year = "2020"
}

@article{Escribano:2023hxj,
    author = "Escribano, Pablo and Lozano, V{\'\i}ctor Mart{\'\i}n and Vicente, Avelino",
    title = "{Scotogenic explanation for the 95~GeV excesses}",
    eprint = "2306.03735",
    archivePrefix = "arXiv",
    primaryClass = "hep-ph",
    reportNumber = "IFIC/23-20",
    doi = "10.1103/PhysRevD.108.115001",
    journal = "Phys. Rev. D",
    volume = "108",
    number = "11",
    pages = "115001",
    year = "2023"
}

@article{Dev:2023kzu,
    author = "Dev, P. S. Bhupal and Mohapatra, Rabindra N. and Zhang, Yongchao",
    title = "{Explanation of the 95 GeV {\ensuremath{\gamma}}{\ensuremath{\gamma}} and bb{\textasciimacron} excesses in the minimal left-right symmetric model}",
    eprint = "2312.17733",
    archivePrefix = "arXiv",
    primaryClass = "hep-ph",
    reportNumber = "CETUP-2023-021",
    doi = "10.1016/j.physletb.2024.138481",
    journal = "Phys. Lett. B",
    volume = "849",
    pages = "138481",
    year = "2024"
}

@article{Ahriche:2023hho,
    author = "Ahriche, Amine and Bellilet, Mohamed Lamine and Khojali, Mohammed Omer and Kumar, Mukesh and Mulaudzi, Anza-Tshildzi",
    title = "{Scale invariant scotogenic model: CDF-II W-boson mass and the 95~GeV excesses}",
    eprint = "2311.08297",
    archivePrefix = "arXiv",
    primaryClass = "hep-ph",
    doi = "10.1103/PhysRevD.110.015025",
    journal = "Phys. Rev. D",
    volume = "110",
    number = "1",
    pages = "015025",
    year = "2024"
}

@article{Biekotter:2019kde,
    author = {Biek{\"o}tter, T. and Chakraborti, M. and Heinemeyer, S.},
    title = "{A 96 GeV Higgs boson in the N2HDM}",
    eprint = "1903.11661",
    archivePrefix = "arXiv",
    primaryClass = "hep-ph",
    reportNumber = "IFT-UAM/CSIC-19-034",
    doi = "10.1140/epjc/s10052-019-7561-2",
    journal = "Eur. Phys. J. C",
    volume = "80",
    number = "1",
    pages = "2",
    year = "2020"
}

@article{Benbrik:2025hol,
    author = "Benbrik, Rachid and Boukidi, Mohammed and Kahime, Khouloud and Moretti, Stefano and Rahili, Larbi and Taki, Bassim",
    title = "{Exploring potential Higgs resonances at 650 GeV and 95 GeV in the 2HDM Type III}",
    eprint = "2505.07811",
    archivePrefix = "arXiv",
    primaryClass = "hep-ph",
    doi = "10.1016/j.physletb.2025.139688",
    journal = "Phys. Lett. B",
    volume = "868",
    pages = "139688",
    year = "2025"
}

@article{Heinemeyer:2021msz,
    author = "Heinemeyer, S. and Li, C. and Lika, F. and Moortgat-Pick, G. and Paasch, S.",
    title = "{Phenomenology of a 96~GeV Higgs boson in the 2HDM with an additional singlet}",
    eprint = "2112.11958",
    archivePrefix = "arXiv",
    primaryClass = "hep-ph",
    reportNumber = "DESY 21-230, IFT-UAM/CSIC-21-158",
    doi = "10.1103/PhysRevD.106.075003",
    journal = "Phys. Rev. D",
    volume = "106",
    number = "7",
    pages = "075003",
    year = "2022"
}

@article{Biekotter:2021ovi,
    author = {Biek{\"o}tter, Thomas and Olea-Romacho, Mar{\'\i}a Olalla},
    title = "{Reconciling Higgs physics and pseudo-Nambu-Goldstone dark matter in the S2HDM using a genetic algorithm}",
    eprint = "2108.10864",
    archivePrefix = "arXiv",
    primaryClass = "hep-ph",
    reportNumber = "DESY-21-125",
    doi = "10.1007/JHEP10(2021)215",
    journal = "JHEP",
    volume = "10",
    pages = "215",
    year = "2021"
}

@article{Banik:2024ugs,
    author = "Banik, Sumit and Coloretti, Guglielmo and Crivellin, Andreas and Haber, Howard E.",
    title = "{Correlating A{\textrightarrow}{\ensuremath{\gamma}}{\ensuremath{\gamma}} with electric dipole moments in the two Higgs doublet model in light of the diphoton excesses at 95~GeV and 152~GeV}",
    eprint = "2412.00523",
    archivePrefix = "arXiv",
    primaryClass = "hep-ph",
    reportNumber = "PSI-PR-24-24, ZU-TH 58/24",
    doi = "10.1103/PhysRevD.111.075021",
    journal = "Phys. Rev. D",
    volume = "111",
    number = "7",
    pages = "075021",
    year = "2025"
}

@article{Dutta:2025nmy,
    author = "Dutta, Juhi and Lahiri, Jayita and Li, Cheng and Moortgat-Pick, Gudrid and Tabira, Sheikh Farah and Ziegler, Julia Anabell",
    title = "{Search for Dark Matter in 2HDMS at LHC and future Lepton Colliders}",
    eprint = "2504.14529",
    archivePrefix = "arXiv",
    primaryClass = "hep-ph",
    reportNumber = "DESY-25-061",
    month = "4",
    year = "2025"
}

@article{Benbrik:2022azi,
    author = "Benbrik, Rachid and Boukidi, Mohammed and Moretti, Stefano and Semlali, Souad",
    title = "{Explaining the 96 GeV Di-photon anomaly in a generic 2HDM Type-III}",
    eprint = "2204.07470",
    archivePrefix = "arXiv",
    primaryClass = "hep-ph",
    doi = "10.1016/j.physletb.2022.137245",
    journal = "Phys. Lett. B",
    volume = "832",
    pages = "137245",
    year = "2022"
}

@article{Biekotter:2020cjs,
    author = {Biek{\"o}tter, T. and Chakraborti, M. and Heinemeyer, S.},
    title = "{The {\textquotedblleft}96 GeV excess{\textquotedblright} at the LHC}",
    eprint = "2003.05422",
    archivePrefix = "arXiv",
    primaryClass = "hep-ph",
    reportNumber = "IFT-UAM/CSIC-20-041, DESY 20-047, DESY-20-047",
    doi = "10.1142/S0217751X21420185",
    journal = "Int. J. Mod. Phys. A",
    volume = "36",
    number = "22",
    pages = "2142018",
    year = "2021"
}

@article{Cao:2016uwt,
    author = "Cao, Junjie and Guo, Xiaofei and He, Yangle and Wu, Peiwen and Zhang, Yang",
    title = "{Diphoton signal of the light Higgs boson in natural NMSSM}",
    eprint = "1612.08522",
    archivePrefix = "arXiv",
    primaryClass = "hep-ph",
    doi = "10.1103/PhysRevD.95.116001",
    journal = "Phys. Rev. D",
    volume = "95",
    number = "11",
    pages = "116001",
    year = "2017"
}

@article{Li:2022etb,
    author = "Li, Weichao and Qiao, Haoxue and Zhu, Jingya",
    title = "{Light Higgs boson in the NMSSM confronted with the CMS di-photon and di-tau excesses*}",
    eprint = "2212.11739",
    archivePrefix = "arXiv",
    primaryClass = "hep-ph",
    doi = "10.1088/1674-1137/acfaf1",
    journal = "Chin. Phys. C",
    volume = "47",
    number = "12",
    pages = "123102",
    year = "2023"
}

@article{Cao:2023gkc,
    author = "Cao, Junjie and Jia, Xinglong and Lian, Jingwei and Meng, Lei",
    title = "{95~GeV diphoton and bb{\textasciimacron} excesses in the general next-to-minimal supersymmetric standard model}",
    eprint = "2310.08436",
    archivePrefix = "arXiv",
    primaryClass = "hep-ph",
    doi = "10.1103/PhysRevD.109.075001",
    journal = "Phys. Rev. D",
    volume = "109",
    number = "7",
    pages = "075001",
    year = "2024"
}

@article{Lian:2024smg,
    author = "Lian, Jingwei",
    title = "{95~GeV excesses in the Z3-symmetric next-to-minimal supersymmetric standard model}",
    eprint = "2406.10969",
    archivePrefix = "arXiv",
    primaryClass = "hep-ph",
    doi = "10.1103/PhysRevD.110.115018",
    journal = "Phys. Rev. D",
    volume = "110",
    number = "11",
    pages = "115018",
    year = "2024"
}

@article{Cao:2024axg,
    author = "Cao, Junjie and Jia, Xinglong and Lian, Jingwei",
    title = "{Unified interpretation of the muon g-2 anomaly, the 95~GeV diphoton, and bb{\textasciimacron} excesses in the general next-to-minimal supersymmetric standard model}",
    eprint = "2402.15847",
    archivePrefix = "arXiv",
    primaryClass = "hep-ph",
    doi = "10.1103/PhysRevD.110.115039",
    journal = "Phys. Rev. D",
    volume = "110",
    number = "11",
    pages = "115039",
    year = "2024"
}

@article{Ellwanger:2023zjc,
    author = "Ellwanger, Ulrich and Hugonie, Cyril",
    title = "{Additional Higgs Bosons near 95 and 650 GeV in the NMSSM}",
    eprint = "2309.07838",
    archivePrefix = "arXiv",
    primaryClass = "hep-ph",
    doi = "10.1140/epjc/s10052-023-12315-y",
    journal = "Eur. Phys. J. C",
    volume = "83",
    number = "12",
    pages = "1138",
    year = "2023"
}

@article{Ellwanger:2024vvs,
    author = "Ellwanger, Ulrich and Hugonie, Cyril and King, Stephen F. and Moretti, Stefano",
    title = "{NMSSM explanation for excesses in the search for neutralinos and charginos and a 95 GeV Higgs boson}",
    eprint = "2404.19338",
    archivePrefix = "arXiv",
    primaryClass = "hep-ph",
    doi = "10.1140/epjc/s10052-024-13129-2",
    journal = "Eur. Phys. J. C",
    volume = "84",
    number = "8",
    pages = "788",
    year = "2024"
}

@article{Choi:2019yrv,
    author = "Choi, Kiwoon and Im, Sang Hui and Jeong, Kwang Sik and Park, Chan Beom",
    title = "{Light Higgs bosons in the general NMSSM}",
    eprint = "1906.03389",
    archivePrefix = "arXiv",
    primaryClass = "hep-ph",
    reportNumber = "CTPU-PTC-19-17, PNUTP-19-A11",
    doi = "10.1140/epjc/s10052-019-7473-1",
    journal = "Eur. Phys. J. C",
    volume = "79",
    number = "11",
    pages = "956",
    year = "2019"
}

@article{Ellwanger:2024txc,
    author = "Ellwanger, Ulrich and Hugonie, Cyril",
    title = "{Nmssm with correct relic density and an additional 95~GeV Higgs boson}",
    eprint = "2403.16884",
    archivePrefix = "arXiv",
    primaryClass = "hep-ph",
    doi = "10.1140/epjc/s10052-024-12886-4",
    journal = "Eur. Phys. J. C",
    volume = "84",
    number = "5",
    pages = "526",
    year = "2024"
}

@article{Chen:2023bqr,
    author = "Chen, Ting-Kuo and Chiang, Cheng-Wei and Heinemeyer, Sven and Weiglein, Georg",
    title = "{95~GeV Higgs boson in the Georgi-Machacek model}",
    eprint = "2312.13239",
    archivePrefix = "arXiv",
    primaryClass = "hep-ph",
    doi = "10.1103/PhysRevD.109.075043",
    journal = "Phys. Rev. D",
    volume = "109",
    number = "7",
    pages = "075043",
    year = "2024"
}

@article{Ge:2024rdr,
    author = "Ge, Zhao-feng and Niu, Feng-Yan and Yang, Jin-Lei",
    title = "{The origin of the 95~GeV excess in the flavor-dependent $U(1)_X$ model}",
    eprint = "2405.07243",
    archivePrefix = "arXiv",
    primaryClass = "hep-ph",
    doi = "10.1140/epjc/s10052-024-12872-w",
    journal = "Eur. Phys. J. C",
    volume = "84",
    number = "5",
    pages = "548",
    year = "2024"
}

@article{YaserAyazi:2024hpj,
    author = "Yaser Ayazi, Seyed and Hosseini, Mojtaba and Paktinat Mehdiabadi, Saeid and Rouzbehi, Rouzbeh",
    title = "{Vector dark matter and LHC constraints, including a 95~GeV light Higgs boson}",
    eprint = "2405.01132",
    archivePrefix = "arXiv",
    primaryClass = "hep-ph",
    doi = "10.1103/PhysRevD.110.055004",
    journal = "Phys. Rev. D",
    volume = "110",
    number = "5",
    pages = "055004",
    year = "2024"
}

@article{Chang:2025bjt,
    author = "Chang, Qin and Du, Xiaokang and Zhu, Pengxuan",
    title = "{Unified interpretation of 95 GeV di-photon and di-tau Excesses in the Georgi-Machacek Model}",
    eprint = "2509.26155",
    archivePrefix = "arXiv",
    primaryClass = "hep-ph",
    month = "9",
    year = "2025"
}

@article{Khanna:2024bah,
    author = "Khanna, Akshat and Moretti, Stefano and Sarkar, Agnivo",
    title = "{Explaining 95 GeV anomalies in the 2-Higgs doublet model type-I}",
    eprint = "2409.02587",
    archivePrefix = "arXiv",
    primaryClass = "hep-ph",
    reportNumber = "HRI-RECAPP-2024-05",
    doi = "10.1016/j.nuclphysb.2025.117229",
    journal = "Nucl. Phys. B",
    volume = "1022",
    pages = "117229",
    year = "2026"
}

@article{Biekotter:2021qbc,
    author = {Biek{\"o}tter, Thomas and Grohsjean, Alexander and Heinemeyer, Sven and Schwanenberger, Christian and Weiglein, Georg},
    title = "{Possible indications for new Higgs bosons in the reach of the LHC: N2HDM and NMSSM interpretations}",
    eprint = "2109.01128",
    archivePrefix = "arXiv",
    primaryClass = "hep-ph",
    reportNumber = "IFT--UAM/CSIC--21-041, IFT-UAM/CSIC-21-041, DESY 21-132",
    doi = "10.1140/epjc/s10052-022-10099-1",
    journal = "Eur. Phys. J. C",
    volume = "82",
    number = "2",
    pages = "178",
    year = "2022"
}

@article{Aguilar-Saavedra:2023tql,
    author = "Aguilar-Saavedra, J. A. and C{\^a}mara, H. B. and Joaquim, F. R. and Seabra, J. F.",
    title = "{Confronting the 95 GeV excesses
within the U(1)'-extended next-to-minimal 2HDM}",
    eprint = "2307.03768",
    archivePrefix = "arXiv",
    primaryClass = "hep-ph",
    reportNumber = "IFT-UAM-CSIC-23-86",
    doi = "10.1103/PhysRevD.108.075020",
    journal = "Phys. Rev. D",
    volume = "108",
    number = "7",
    pages = "075020",
    year = "2023"
}

@article{Biekotter:2022jyr,
    author = {Biek{\"o}tter, Thomas and Heinemeyer, Sven and Weiglein, Georg},
    title = "{Mounting evidence for a 95 GeV Higgs boson}",
    eprint = "2203.13180",
    archivePrefix = "arXiv",
    primaryClass = "hep-ph",
    reportNumber = "DESY 22-057, IFT-UAM/CSIC-22-033, IFT-UAM/CSIC--22--033",
    doi = "10.1007/JHEP08(2022)201",
    journal = "JHEP",
    volume = "08",
    pages = "201",
    year = "2022"
}

@article{Biekotter:2022abc,
    author = {Biek{\"o}tter, Thomas and Heinemeyer, Sven and Weiglein, Georg},
    title = "{Excesses in the low-mass Higgs-boson search and the $\varvec{W}$-boson mass measurement}",
    eprint = "2204.05975",
    archivePrefix = "arXiv",
    primaryClass = "hep-ph",
    reportNumber = "DESY 22-067, IFT-UAM/CSIC--22--043",
    doi = "10.1140/epjc/s10052-023-11635-3",
    journal = "Eur. Phys. J. C",
    volume = "83",
    number = "5",
    pages = "450",
    year = "2023"
}

@article{Biekotter:2023jld,
    author = {Biek{\"o}tter, Thomas and Heinemeyer, Sven and Weiglein, Georg},
    title = "{The CMS di-photon excess at 95 GeV in view of the LHC Run 2 results}",
    eprint = "2303.12018",
    archivePrefix = "arXiv",
    primaryClass = "hep-ph",
    reportNumber = "KA-TP-03-2023, DESY-23-033, IFT-UAM/CSIC-23-028",
    doi = "10.1016/j.physletb.2023.138217",
    journal = "Phys. Lett. B",
    volume = "846",
    pages = "138217",
    year = "2023"
}

@article{Biekotter:2023oen,
    author = {Biek{\"o}tter, Thomas and Heinemeyer, Sven and Weiglein, Georg},
    title = "{95.4~GeV diphoton excess at ATLAS and CMS}",
    eprint = "2306.03889",
    archivePrefix = "arXiv",
    primaryClass = "hep-ph",
    reportNumber = "KA-TP-11-2023, DESY-23-071, IFT--UAM/CSIC-23-062",
    doi = "10.1103/PhysRevD.109.035005",
    journal = "Phys. Rev. D",
    volume = "109",
    number = "3",
    pages = "035005",
    year = "2024"
}

@article{Arcadi:2023smv,
    author = "Arcadi, Giorgio and Busoni, Giorgio and Cabo-Almeida, David and Krishnan, Navneet",
    title = "{Is there a scalar or pseudoscalar at 95~GeV?}",
    eprint = "2311.14486",
    archivePrefix = "arXiv",
    primaryClass = "hep-ph",
    doi = "10.1103/PhysRevD.110.115028",
    journal = "Phys. Rev. D",
    volume = "110",
    number = "11",
    pages = "115028",
    year = "2024"
}

@article{Benbrik:2024ptw,
    author = "Benbrik, Rachid and Boukidi, Mohammed and Moretti, Stefano",
    title = "{Superposition of CP-even and CP-odd Higgs resonances: Explaining the 95~GeV excesses within a two-Higgs-doublet model}",
    eprint = "2405.02899",
    archivePrefix = "arXiv",
    primaryClass = "hep-ph",
    doi = "10.1103/PhysRevD.110.115030",
    journal = "Phys. Rev. D",
    volume = "110",
    number = "11",
    pages = "115030",
    year = "2024"
}

@article{Iguro:2022fel,
    author = "Iguro, Syuhei and Kitahara, Teppei and Omura, Yuji and Zhang, Hantian",
    title = "{Chasing the two-Higgs doublet model in the di-Higgs boson production}",
    eprint = "2211.00011",
    archivePrefix = "arXiv",
    primaryClass = "hep-ph",
    reportNumber = "P3H-22-102, TTP22-061, KEK-TH-2468",
    doi = "10.1103/PhysRevD.107.075017",
    journal = "Phys. Rev. D",
    volume = "107",
    number = "7",
    pages = "075017",
    year = "2023"
}

@article{Azevedo:2023zkg,
    author = {Azevedo, Duarte and Biek{\"o}tter, Thomas and Ferreira, P. M.},
    title = "{2HDM interpretations of the CMS diphoton excess at 95 GeV}",
    eprint = "2305.19716",
    archivePrefix = "arXiv",
    primaryClass = "hep-ph",
    reportNumber = "KA-TP-10-2023",
    doi = "10.1007/JHEP11(2023)017",
    journal = "JHEP",
    volume = "11",
    pages = "017",
    year = "2023"
}

@article{Belyaev:2023xnv,
    author = "Belyaev, Alexander and Benbrik, Rachid and Boukidi, Mohammed and Chakraborti, Manimala and Moretti, Stefano and Semlali, Souad",
    title = "{Explanation of the hints for a 95 GeV Higgs boson within a 2-Higgs Doublet Model}",
    eprint = "2306.09029",
    archivePrefix = "arXiv",
    primaryClass = "hep-ph",
    doi = "10.1007/JHEP05(2024)209",
    journal = "JHEP",
    volume = "05",
    pages = "209",
    year = "2024"
}

@article{Mondal:2025tzi,
    author = "Mondal, Poulami and Samanta, Subrata",
    title = "{Light Scalars in the Extended Georgi-Machacek Model}",
    eprint = "2506.06427",
    archivePrefix = "arXiv",
    primaryClass = "hep-ph",
    month = "6",
    year = "2025"
}

@article{Kundu:2024sip,
    author = "Kundu, Anirban and Mondal, Poulami and Moultaka, Gilbert",
    title = "{Indications for new scalar resonances at the LHC and a possible interpretation}",
    eprint = "2411.14126",
    archivePrefix = "arXiv",
    primaryClass = "hep-ph",
    month = "11",
    year = "2024"
}

@article{Hmissou:2025uep,
    author = "Hmissou, Ayoub and Moretti, Stefano and Rahili, Larbi",
    title = "{Investigating the 95~GeV Higgs boson excesses within the type-I (1+2)HDM}",
    eprint = "2502.03631",
    archivePrefix = "arXiv",
    primaryClass = "hep-ph",
    doi = "10.1103/f571-lmgf",
    journal = "Phys. Rev. D",
    volume = "113",
    number = "1",
    pages = "015024",
    year = "2026"
}

@article{Hmissou:2025riw,
    author = "Hmissou, Ayoub and Moretti, Stefano and Rahili, Larbi",
    title = "{Could the 650~GeV excess be a pseudoscalar of a three-Higgs-doublet model?}",
    eprint = "2509.06232",
    archivePrefix = "arXiv",
    primaryClass = "hep-ph",
    doi = "10.1103/6qs7-869p",
    journal = "Phys. Rev. D",
    volume = "112",
    number = "7",
    pages = "075049",
    year = "2025"
}

@article{Benbrik:2025wkz,
    author = "Benbrik, Rachid and Boukidi, Mohammed and Kahime, Khouloud and Moretti, Stefano and Rahili, Larbi and Taki, Bassim",
    title = "{Interpreting the 650 GeV and 95 GeV Higgs Anomalies in the N2HDM}",
    eprint = "2510.19605",
    archivePrefix = "arXiv",
    primaryClass = "hep-ph",
    month = "10",
    year = "2025"
}

@article{Sassi:2025dyj,
    author = "Sassi, Mohamed Younes and Moortgat-Pick, Gudrid",
    title = "{Vacuum instability and false vacuum decay induced by domain walls in the N2HDM}",
    eprint = "2506.14880",
    archivePrefix = "arXiv",
    primaryClass = "hep-ph",
    reportNumber = "DESY-25-084",
    doi = "10.1140/epjc/s10052-025-14875-7",
    journal = "Eur. Phys. J. C",
    volume = "85",
    number = "10",
    pages = "1230",
    year = "2025"
}

@article{Chen:2013jvg,
    author = "Chen, Chien-Yi and Freid, Michael and Sher, Marc",
    title = "{Next-to-minimal two Higgs doublet model}",
    eprint = "1312.3949",
    archivePrefix = "arXiv",
    primaryClass = "hep-ph",
    reportNumber = "JLAB-THY-14-1839",
    doi = "10.1103/PhysRevD.89.075009",
    journal = "Phys. Rev. D",
    volume = "89",
    number = "7",
    pages = "075009",
    year = "2014"
}

@article{Drozd:2014yla,
    author = "Drozd, Aleksandra and Grzadkowski, Bohdan and Gunion, John F. and Jiang, Yun",
    title = "{Extending two-Higgs-doublet models by a singlet scalar field - the Case for Dark Matter}",
    eprint = "1408.2106",
    archivePrefix = "arXiv",
    primaryClass = "hep-ph",
    doi = "10.1007/JHEP11(2014)105",
    journal = "JHEP",
    volume = "11",
    pages = "105",
    year = "2014"
}

@article{vonBuddenbrock:2016rmr,
    author = "von Buddenbrock, Stefan and Chakrabarty, Nabarun and Cornell, Alan S. and Kar, Deepak and Kumar, Mukesh and Mandal, Tanumoy and Mellado, Bruce and Mukhopadhyaya, Biswarup and Reed, Robert G. and Ruan, Xifeng",
    title = "{Phenomenological signatures of additional scalar bosons at the LHC}",
    eprint = "1606.01674",
    archivePrefix = "arXiv",
    primaryClass = "hep-ph",
    reportNumber = "WITS-MITP-025, HRI-RECAPP-2016-003",
    doi = "10.1140/epjc/s10052-016-4435-8",
    journal = "Eur. Phys. J. C",
    volume = "76",
    number = "10",
    pages = "580",
    year = "2016"
}

@article{Muhlleitner:2016mzt,
    author = "Muhlleitner, Margarete and Sampaio, Marco O. P. and Santos, Rui and Wittbrodt, Jonas",
    title = "{The N2HDM under Theoretical and Experimental Scrutiny}",
    eprint = "1612.01309",
    archivePrefix = "arXiv",
    primaryClass = "hep-ph",
    doi = "10.1007/JHEP03(2017)094",
    journal = "JHEP",
    volume = "03",
    pages = "094",
    year = "2017"
}

@article{Coimbra:2013qq,
    author = "Coimbra, Rita and Sampaio, Marco O. P. and Santos, Rui",
    title = "{ScannerS: Constraining the phase diagram of a complex scalar singlet at the LHC}",
    eprint = "1301.2599",
    archivePrefix = "arXiv",
    primaryClass = "hep-ph",
    doi = "10.1140/epjc/s10052-013-2428-4",
    journal = "Eur. Phys. J. C",
    volume = "73",
    pages = "2428",
    year = "2013"
}

@article{Muhlleitner:2020wwk,
    author = {M{\"u}hlleitner, Margarete and Sampaio, Marco O. P. and Santos, Rui and Wittbrodt, Jonas},
    title = "{ScannerS: parameter scans in extended scalar sectors}",
    eprint = "2007.02985",
    archivePrefix = "arXiv",
    primaryClass = "hep-ph",
    reportNumber = "KA-TP-05-2020, LU TP 20-38",
    doi = "10.1140/epjc/s10052-022-10139-w",
    journal = "Eur. Phys. J. C",
    volume = "82",
    number = "3",
    pages = "198",
    year = "2022"
}

@article{Kanemura:1993hm,
    author = "Kanemura, Shinya and Kubota, Takahiro and Takasugi, Eiichi",
    title = "{Lee-Quigg-Thacker bounds for Higgs boson masses in a two doublet model}",
    eprint = "hep-ph/9303263",
    archivePrefix = "arXiv",
    reportNumber = "OS-GE-32-93",
    doi = "10.1016/0370-2693(93)91205-2",
    journal = "Phys. Lett. B",
    volume = "313",
    pages = "155--160",
    year = "1993"
}

@article{Hollik:2018wrr,
    author = "Hollik, Wolfgang G. and Weiglein, Georg and Wittbrodt, Jonas",
    title = "{Impact of Vacuum Stability Constraints on the Phenomenology of Supersymmetric Models}",
    eprint = "1812.04644",
    archivePrefix = "arXiv",
    primaryClass = "hep-ph",
    reportNumber = "DESY 18-148, DESY-18-148, TTP 18-036",
    doi = "10.1007/JHEP03(2019)109",
    journal = "JHEP",
    volume = "03",
    pages = "109",
    year = "2019"
}

@article{Ferreira:2019iqb,
    author = {Ferreira, P. M. and M{\"u}hlleitner, Margarete and Santos, Rui and Weiglein, Georg and Wittbrodt, Jonas},
    title = "{Vacuum Instabilities in the N2HDM}",
    eprint = "1905.10234",
    archivePrefix = "arXiv",
    primaryClass = "hep-ph",
    reportNumber = "DESY-19-085, DESY 19-085, KA-TP-07-2019",
    doi = "10.1007/JHEP09(2019)006",
    journal = "JHEP",
    volume = "09",
    pages = "006",
    year = "2019"
}

@article{Haller:2018nnx,
    author = {Haller, Johannes and Hoecker, Andreas and Kogler, Roman and M{\"o}nig, Klaus and Peiffer, Thomas and Stelzer, J{\"o}rg},
    title = "{Update of the global electroweak fit and constraints on two-Higgs-doublet models}",
    eprint = "1803.01853",
    archivePrefix = "arXiv",
    primaryClass = "hep-ph",
    doi = "10.1140/epjc/s10052-018-6131-3",
    journal = "Eur. Phys. J. C",
    volume = "78",
    number = "8",
    pages = "675",
    year = "2018"
}

@article{Bechtle:2020pkv,
    author = "Bechtle, Philip and Dercks, Daniel and Heinemeyer, Sven and Klingl, Tobias and Stefaniak, Tim and Weiglein, Georg and Wittbrodt, Jonas",
    title = "{HiggsBounds-5: Testing Higgs Sectors in the LHC 13 TeV Era}",
    eprint = "2006.06007",
    archivePrefix = "arXiv",
    primaryClass = "hep-ph",
    reportNumber = "BONN-TH-2020-03, DESY 20-093, DESY-20-093, IFT-UAM/CSIC-20-072, LU 20-27",
    doi = "10.1140/epjc/s10052-020-08557-9",
    journal = "Eur. Phys. J. C",
    volume = "80",
    number = "12",
    pages = "1211",
    year = "2020"
}

@article{Bechtle:2020uwn,
    author = "Bechtle, Philip and Heinemeyer, Sven and Klingl, Tobias and Stefaniak, Tim and Weiglein, Georg and Wittbrodt, Jonas",
    title = "{HiggsSignals-2: Probing new physics with precision Higgs measurements in the LHC 13 TeV era}",
    eprint = "2012.09197",
    archivePrefix = "arXiv",
    primaryClass = "hep-ph",
    reportNumber = "BONN-TH-2020-09, DESY-20-228, DESY 20-228, IFT-UAM/CSIC-20-081, LU TP 20-53",
    doi = "10.1140/epjc/s10052-021-08942-y",
    journal = "Eur. Phys. J. C",
    volume = "81",
    number = "2",
    pages = "145",
    year = "2021"
}

@article{Engeln:2018mbg,
    author = {Engeln, Isabell and M{\"u}hlleitner, Margarete and Wittbrodt, Jonas},
    title = "{N2HDECAY: Higgs Boson Decays in the Different Phases of the N2HDM}",
    eprint = "1805.00966",
    archivePrefix = "arXiv",
    primaryClass = "hep-ph",
    reportNumber = "DESY 18-064, KA-TP-09-2018, DESY-18-064",
    doi = "10.1016/j.cpc.2018.07.020",
    journal = "Comput. Phys. Commun.",
    volume = "234",
    pages = "256--262",
    year = "2019"
}

@article{ParticleDataGroup:2024cfk,
    author = "Navas, S. and others",
    collaboration = "Particle Data Group",
    title = "{Review of particle physics}",
    doi = "10.1103/PhysRevD.110.030001",
    journal = "Phys. Rev. D",
    volume = "110",
    number = "3",
    pages = "030001",
    year = "2024"
}

@article{Chen:2016zpw,
    author = "Chen, Zhenxing and Yang, Ying and Ruan, Manqi and Wang, Dayong and Li, Gang and Jin, Shan and Ban, Yong",
    title = "{Cross Section and Higgs Mass Measurement with Higgsstrahlung at the CEPC}",
    eprint = "1601.05352",
    archivePrefix = "arXiv",
    primaryClass = "hep-ex",
    doi = "10.1088/1674-1137/41/2/023003",
    journal = "Chin. Phys. C",
    volume = "41",
    number = "2",
    pages = "023003",
    year = "2017"
}

@article{Yang:2022qga,
    author = "Yang, Jin Min and Zhang, Yang and Zhu, Pengxuan and Zhu, Rui",
    title = "{Reconstructing masses for semi-invisibly decaying particles pair-produced at lepton colliders}",
    eprint = "2211.08132",
    archivePrefix = "arXiv",
    primaryClass = "hep-ph",
    doi = "10.1103/PhysRevD.108.075015",
    journal = "Phys. Rev. D",
    volume = "108",
    number = "7",
    pages = "075015",
    year = "2023"
}

@article{Alwall:2011uj,
    author = "Alwall, Johan and Herquet, Michel and Maltoni, Fabio and Mattelaer, Olivier and Stelzer, Tim",
    title = "{MadGraph 5 : Going Beyond}",
    eprint = "1106.0522",
    archivePrefix = "arXiv",
    primaryClass = "hep-ph",
    reportNumber = "FERMILAB-PUB-11-448-T",
    doi = "10.1007/JHEP06(2011)128",
    journal = "JHEP",
    volume = "06",
    pages = "128",
    year = "2011"
}

@article{Alwall:2014hca,
    author = "Alwall, J. and Frederix, R. and Frixione, S. and Hirschi, V. and Maltoni, F. and Mattelaer, O. and Shao, H. -S. and Stelzer, T. and Torrielli, P. and Zaro, M.",
    title = "{The automated computation of tree-level and next-to-leading order differential cross sections, and their matching to parton shower simulations}",
    eprint = "1405.0301",
    archivePrefix = "arXiv",
    primaryClass = "hep-ph",
    reportNumber = "CERN-PH-TH-2014-064, CP3-14-18, LPN14-066, MCNET-14-09, ZU-TH-14-14",
    doi = "10.1007/JHEP07(2014)079",
    journal = "JHEP",
    volume = "07",
    pages = "079",
    year = "2014"
}

@article{Cacciari:2008gp,
    author = "Cacciari, Matteo and Salam, Gavin P. and Soyez, Gregory",
    title = "{The anti-$k_t$ jet clustering algorithm}",
    eprint = "0802.1189",
    archivePrefix = "arXiv",
    primaryClass = "hep-ph",
    reportNumber = "LPTHE-07-03",
    doi = "10.1088/1126-6708/2008/04/063",
    journal = "JHEP",
    volume = "04",
    pages = "063",
    year = "2008"
}

@article{Yu:2020bxh,
    author = "Yu, Dan and Ruan, Manqi and Boudry, Vincent and Videau, Henri and Brient, Jean-Claude and Wu, Zhigang and Ouyang, Qun and Xu, Yue and Chen, Xin",
    title = "{The measurement of the $H\rightarrow \tau \tau $ signal strength in the future $e^{+}e^{-}$ Higgs factories}",
    doi = "10.1140/epjc/s10052-019-7557-y",
    journal = "Eur. Phys. J. C",
    volume = "80",
    number = "1",
    pages = "7",
    year = "2020"
}

@article{Sjostrand:2014zea,
    author = {Sj{\"o}strand, Torbj{\"o}rn and Ask, Stefan and Christiansen, Jesper R. and Corke, Richard and Desai, Nishita and Ilten, Philip and Mrenna, Stephen and Prestel, Stefan and Rasmussen, Christine O. and Skands, Peter Z.},
    title = "{An introduction to PYTHIA 8.2}",
    eprint = "1410.3012",
    archivePrefix = "arXiv",
    primaryClass = "hep-ph",
    reportNumber = "LU-TP-14-36, MCNET-14-22, CERN-PH-TH-2014-190, FERMILAB-PUB-14-316-CD, DESY-14-178, SLAC-PUB-16122",
    doi = "10.1016/j.cpc.2015.01.024",
    journal = "Comput. Phys. Commun.",
    volume = "191",
    pages = "159--177",
    year = "2015"
}

@article{deFavereau:2013fsa,
    author = "de Favereau, J. and Delaere, C. and Demin, P. and Giammanco, A. and Lema{\^\i}tre, V. and Mertens, A. and Selvaggi, M.",
    collaboration = "DELPHES 3",
    title = "{DELPHES 3, A modular framework for fast simulation of a generic collider experiment}",
    eprint = "1307.6346",
    archivePrefix = "arXiv",
    primaryClass = "hep-ex",
    doi = "10.1007/JHEP02(2014)057",
    journal = "JHEP",
    volume = "02",
    pages = "057",
    year = "2014"
}

@article{Selvaggi:2014mya,
    author = "Selvaggi, Michele",
    editor = "Wang, Jianxiong",
    title = "{DELPHES 3: A modular framework for fast-simulation of generic collider experiments}",
    doi = "10.1088/1742-6596/523/1/012033",
    journal = "J. Phys. Conf. Ser.",
    volume = "523",
    pages = "012033",
    year = "2014"
}

@article{Zheng:2020ult,
    author = "Zheng, Taifan and Xu, Ji and Cao, Lu and Yu, Dan and Wang, Wei and Prell, Soeren and Cheung, Yeuk-Kwan E. and Ruan, Manqi",
    title = "{Analysis of $B_c \to \tau\nu_\tau$ at CEPC}",
    eprint = "2007.08234",
    archivePrefix = "arXiv",
    primaryClass = "hep-ex",
    doi = "10.1088/1674-1137/abcf1f",
    journal = "Chin. Phys. C",
    volume = "45",
    number = "2",
    pages = "023001",
    year = "2021"
}

@article{Zhu:2025eoe,
    author = "Zhu, Yongfeng and Liang, Hao and Wang, Yuexin and Che, Yuzhi and Wang, Hengyu and Zhou, Chen and Qu, Huilin and Ruan, Manqi",
    title = "{Holistic approach and Advanced Color Singlet Identification for physics measurements at high energy frontier}",
    eprint = "2506.11783",
    archivePrefix = "arXiv",
    primaryClass = "hep-ex",
    month = "6",
    year = "2025"
}

@article{Cowan:2010js,
    author = "Cowan, Glen and Cranmer, Kyle and Gross, Eilam and Vitells, Ofer",
    title = "{Asymptotic formulae for likelihood-based tests of new physics}",
    eprint = "1007.1727",
    archivePrefix = "arXiv",
    primaryClass = "physics.data-an",
    doi = "10.1140/epjc/s10052-011-1554-0",
    journal = "Eur. Phys. J. C",
    volume = "71",
    pages = "1554",
    year = "2011",
    note = "[Erratum: Eur.Phys.J.C 73, 2501 (2013)]"
}

@article{Du:2025eop,
    author = "Du, Xiaokang and Liu, Huiling and Chang, Qin",
    title = "{Interpretation of 95~GeV excess within the Georgi-Machacek model in light of positive definiteness constraints}",
    eprint = "2502.06444",
    archivePrefix = "arXiv",
    primaryClass = "hep-ph",
    doi = "10.1103/w2xw-q49n",
    journal = "Phys. Rev. D",
    volume = "112",
    number = "1",
    pages = "015019",
    year = "2025"
}

@article{Ahriche:2023wkj,
    author = "Ahriche, Amine",
    title = "{95~GeV excess in the Georgi-Machacek model: Single or twin peak resonance}",
    eprint = "2312.10484",
    archivePrefix = "arXiv",
    primaryClass = "hep-ph",
    doi = "10.1103/PhysRevD.110.035010",
    journal = "Phys. Rev. D",
    volume = "110",
    number = "3",
    pages = "035010",
    year = "2024"
}

@article{CEPCStudyGroup:2025kmw,
    author = "Adhya, Souvik Priyam and others",
    collaboration = "CEPC Study Group",
    title = "{CEPC Technical Design Report - Reference Detector}",
    eprint = "2510.05260",
    archivePrefix = "arXiv",
    primaryClass = "hep-ex",
    reportNumber = "IHEP-CEPC-DR-2025-01, IHEP-EP-2025-01",
    month = "10",
    year = "2025"
}

\end{document}